\documentclass[lettersize,journal]{IEEEtran}
\usepackage{amsmath,amsfonts}
\usepackage{algorithm}
\usepackage{algorithmic}
\usepackage{array}
\usepackage[caption=false,font=normalsize,labelfont=sf,textfont=sf]{subfig}
\usepackage{textcomp}
\usepackage{stfloats}
\usepackage{caption}
\usepackage{float}
\usepackage{multirow}
\usepackage{color}
\usepackage{url}
\usepackage{verbatim}
\usepackage{graphicx}
\usepackage{mathrsfs} 
\def\BibTeX{{\rm B\kern-.05em{\sc i\kern-.025em b}\kern-.08em
    T\kern-.1667em\lower.7ex\hbox{E}\kern-.125emX}}
\usepackage{balance}

\begin{document}
\title{Terahertz Integrated Sensing Communications and Powering for 6G Wireless Networks}
\author{
Hua Yan, \textit{Member, IEEE}, Yunfei Chen, \textit{Senior Member, IEEE}
\thanks{H. Yan is with the Department of Engineering, University of Durham, Durham, DH1 3LE, U.K. (e-mail: hua.yan@durham.ac.uk)}
\thanks{Y. Chen is with the Department of Engineering, University of Durham, Durham, DH1 3LE, U.K. (e-mail: yunfei.chen@durham.ac.uk)}
}
\markboth{IEEE Transactions on xxx xxx,~Vol.~xx, No.~x, December~2024}
{Hua, Yunfei, \MakeLowercase{`text{(et al.)}}: THz Integrated Sensing Communications and Powering for 6G Wireless Networks}
\maketitle

\begin{abstract}
\label{Abstract}
The terahertz (THz) band has attracted significant interest for future wireless networks. In this paper, a THz integrated sensing communications and powering (THz-ISCAP) system, where sensing is leveraged to enhance communications and powering, is studied. For a given total amount of time, we aim to determine an optimal time allocation for sensing to improve the efficiency of communications and powering, along with an optimal power splitting ratio to balance these two functionalities. This is achieved by maximizing between communications and powering either the achievable rate or harvested energy while ensuring a minimum requirement on the other. Numerical results indicate that the optimal system performance can be achieved by jointly optimizing the sensing time allocation and the power splitting ratio. Additionally, the results reveal the effects of various factors, such as THz frequencies and antenna aperture sizes, on the system performance. This study provides some interesting results to offer a new perspective for the research on THz-ISCAP.
\end{abstract}


\section{Introduction}
\label{Introduction}
\noindent \IEEEPARstart{R}{ecently}, the terahertz (THz) band (i.e., frequencies spanning from 100 GHz to 10 THz) has attracted widespread attention from both academia and industry, and it has been being studied as a promising technology for the upcoming six-generation (6G) wireless networks \cite{J.M.Jornet2024, W.Jiang2024}. Benefiting from its wide bandwidth, the THz band is able to support and meet the ultra-high data rate requirements of 6G wireless communications. On the other hand, due to the fact that the wavelength of THz signals is less than one millimeter, it has a higher space resolution, leading to high-precision THz radar in target detection. In addition, the THz waves have a very narrow beam width, which can efficiently concentrate energy and supply power to small-scale Internet of Things (IoT) devices \cite{S.Mizojiri2018, Y.Pan2022}. Because of these benefits, many interesting research directions have been explored, such as THz integrated sensing and communications (THz-ISAC) \cite{J.M.Jornet2024, W.Jiang2024}, THz simultaneous wireless information and power transfer (THz-SWIPT) \cite{C.Psomas2024} and THz wireless power transfer (THz-WPT) \cite{S.Mizojiri2018, S.Mizojiri2019}. 

\subsubsection{THz-ISAC}
In \cite{M.Usman2022}, a THz-ISAC system for plant health monitoring in precision agriculture has been discussed as a 6G use case. 
In \cite{G.Wang2022}, THz-ISAC has been discussed and examined from aspects of application scenarios, channel modeling, hardware and measurement campaigns. 
In \cite{Z.Wang2023}, THz-ISAC has been investigated from the aspects of near-field, and some new characteristics, such as spherical wave, have been discussed.
The authors in \cite{Z.Liu2024} investigated the coexistence, coordination and cooperation of THz-ISAC from a theoretical perspective. This work provided metrics for performance analysis of THz-ISAC.
In \cite{T.Gan2024}, the THz-ISAC system architecture and its evolutionary stages and integration modes were discussed, based on which the authors also envisioned the potential research directions from perspectives of ubiquity, intelligence and security. 
In \cite{C.Han2024}, channel modeling and signal processing techniques have been discussed to improve the performance of THz-ISAC.
In \cite{F.Gao2023}, a novel THz-ISAC technique for massive multi-input multi-output (MIMO) has been proposed, where the beam-squint and beam-split were studied to calculate the sensing directions and sensing range, respectively.
In \cite{W.Chen2024}, an ISAC-based beam alignment approach has been proposed to improve the coverage probability of the THz-ISAC network.
In \cite{Y.Wu2024}, a system framework has been proposed to analyze the coverage probability and capacity in THz-ISAC networks.
In \cite{A.M.Elbir2024}, technologies for THz-ISAC, such as antenna and array design, hybrid beam forming, have been examined to discuss the challenges and possible future research directions. 

\subsubsection{THz-SWIPT}
In \cite{Y.Pan2022}, a reconfigurable intelligent surface (RIS) assisted simultaneous THz information and power transfer has been studied. In their work, the information users’ sum rate has been maximized while meeting the energy users’ and RIS’s energy requirements. To characterize the achievable rate and harvested energy of THz-SWIPT systems, the authors in \cite{Z.Yang2023} investigated a THz-SWIPT system employing ultra-massive MIMO with a power splitting structure. Using electromagnetic analysis, the authors derived an approximate closed-form expression for energy distribution. In \cite{N.Shanin2024}, the information rate-harvested power tradeoff in THz-SWIPT systems employing resonant tunneling diode-based (RTD) energy harvesting circuit has been firstly investigated, where all non-linear non-monotonic characteristics of the RTD-based receiver circuit have been taken into account. Besides, reference \cite{N.Shanin2024} also proposed a general non-linear non-monotonic energy harvesting model, which is very helpful for analyzing the performance of the THz-SWIPT systems.

\subsubsection{THz-WPT}
WPT via sub-THz wave was first studied in \cite{S.Mizojiri2018, S.Mizojiri2019}. They concluded that the beam efficiency increases with the frequency, and THz-WPT is suitable for long-distance high-power wireless energy transfer. However, the radio frequency to direct current (RF-to-DC) efficiency decreases with the frequency. In \cite{N.Shanin2023}, a THz-WPT system where an energy harvesting circuit employing a RTD was adopted at the receiver has been studied, through which a general non-linear piecewise energy harvesting model was proposed. 

All these works have provided very useful insights on the study of THz band in 6G wireless networks. To envision the future 6G wireless networks, integrated sensing communications and powering (ISCAP) has been expected to further enhance the usage efficiency of the radio resources \cite{YLChen2024, X.Li2024, Z.Zhou2024}. However, to the best of the authors' knowledge, THz-ISCAP has not been studied yet except for the work in \cite{A.Hanif2023} where THz imaging with information and power transfer was studied. In \cite{A.Hanif2023}, imaging can be considered as a form of sensing. However, sensing also includes positioning, spectroscopy, as well as target detection and localization \cite{W.Jiang2024}. Further, the users of sensing, communications, and powering in ISCAP can be the same or separate. For example, the users studied in \cite{YLChen2024, X.Li2024, Z.Zhou2024} were separate, where the sensing target, the information receiver and the energy receiver are separate. In \cite{A.Hanif2023}, there is only one receiver performing imaging, communications and powering at the same time. As a result, how to apply ISCAP to future communications scenarios is an issue for consideration. Particularly, in \cite{B.Chang2022}, sensing has been used to monitor and evaluate the state of an unmanned aerial vehicle (UAV) to improve the beam alignment between the BS and the UAV for the purpose of communications. This is a typical scenario of sensing-assisted THz communications. The authors in \cite{YLChen2024, X.Li2024, Z.Zhou2024} have worked on ISCAP, but they did not use the THz band. THz bands have distinct characteristics compared with other lower frequency bands, such as small wavelength, atmospheric or molecular absorption, pencil beam \cite{W.Jiang2024} and hence, it is interesting to study ISCAP in the THz band.

Motivated by the above observations, in this paper, we study a THz-ISCAP system where sensing is used to adjust and improve the beam alignment between the transmitter (Tx) and the receiver (Rx), thereby enhancing the performances of communications and powering. For a fixed time frame, the proposed THz-ISCAP system works in the following two stages. In the first stage, similar to \cite{B.Chang2022}, the beam alignment is adjusted through sensing. In the second stage, the remaining allocated time is utilized for SWIPT from the Tx to the Rx. As a result, there exists a time allocation problem between the two stages, and also a performance trade-off between communications and powering in the second stage. The objective is to maximize the received energy while meeting a required data rate or maximize the achievable data rate under energy demand conditions. Simulation results show that there is an optimal time allocation ratio between the two stages, and that there is also an optimal power splitting ratio between communications and powering in the second stage. The main contributions of this paper can be summarized as follows:
\begin{itemize}
    \item A THz-ISCAP system where alignment between the Tx and the Rx is adjusted through sensing to improve the performances of communications and powering is proposed and studied for possible applications in future 6G wireless networks.
    \item To maximize the communications and powering performances, the optimal sensing time ratio and optimal power splitting ratio are obtained.
    \item The effect of aperture size and THz frequency on the system performance are examined for system designs.
\end{itemize}

The rest of this paper is organized as follows. In Section \ref{System Model}, system models are presented. In Section \ref{PF and Solutions}, the problem formulation is established and solved, followed by simulation results and discussion in Section \ref{Numerical Results}. Finally, the paper is concluded in Section \ref{Conclusions}.

\section{System Model}
\label{System Model}
\begin{figure}[!t]
\centering
\includegraphics[width=1\linewidth]{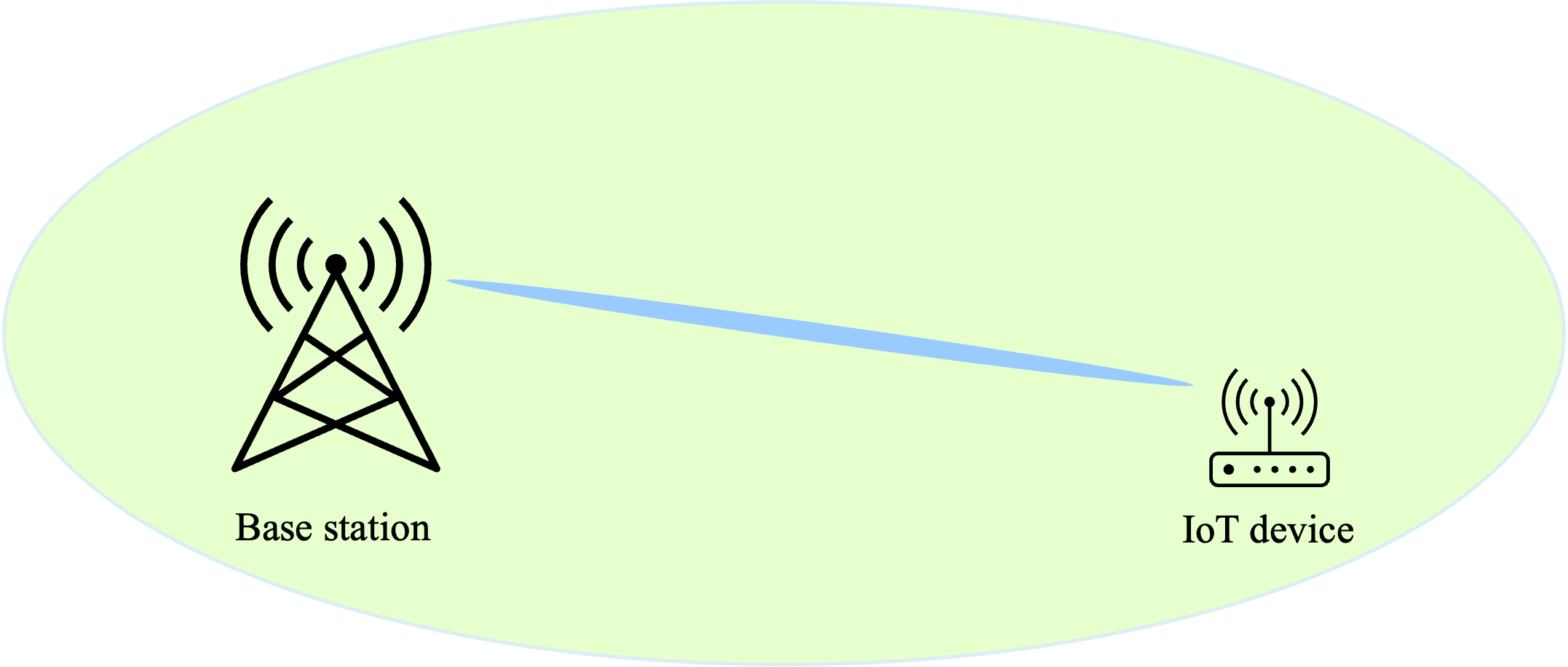}
\caption{System diagram for THz integrated sensing, communications and powering.}
\label{fig1-system-design}
\end{figure}
Consider a THz-ISCAP scenario as depicted in Fig. \ref{fig1-system-design}, where a multi-functional base station (BS) with a THz transmitter sends a THz signal towards a IoT device, i.e., a sensor or a micro drone. The proposed THz-ISCAP scheme can be described as follows. First, the BS sends a THz sensing signal to the receiver, and the signal is reflected and returned to the BS for adjusting the alignment. Then, the BS sends THz signals to the receiver for simultaneous wireless information and power transfer (SWIPT). The total time interval is $T$. The goal is to derive the optimal ratio of sensing and SWIPT time allocation, as well as the optimal power ratio between wireless information and powering. To evaluate and balance the performance of the proposed THz-ISCAP, it is necessary to model the THz signal delivery characteristics for sensing, communications and powering.

\subsection{THz Signal Propagation}
\subsubsection{Antenna Gain}
From \cite{IEEE-standard-2014} and \cite{Hrovat2024}, the aperture antenna gain is determined by the effective area of the antenna and the wavelength, and it is given by
\begin{equation}
\label{eqn-AntennaGain}
G = \eta_{a} \dfrac{4\pi A_{e}}{\lambda^2},
\end{equation}
where $\eta_{a}$ is the aperture efficiency, $A_{e} = \pi({D}/{2})^2$ is the effective area of the antenna, $D$ is the aperture diameter, $\lambda={c}/{f}$ is the wavelength, $c$ is the speed of light, $f$ is the operating frequency at THz. Denote $D_{bs-t}$, $D_{bs-r}$ and $D_{r}$ as the aperture diameters of the BS transmitting antenna, BS receiving antenna and the user receiver antenna. Their effective gains can be derived respectively using \eqref{eqn-AntennaGain} as
\begin{eqnarray}
\label{eqns-AntennaGains}
G_{bs-t} &=& \eta_{bs-t} \left(\dfrac{\pi D_{bs-t}}{\lambda}\right)^2, \\
G_{bs-r} &=& \eta_{bs-r} \left(\dfrac{\pi D_{bs-r}}{\lambda}\right)^2, \\
G_{r} &=& \eta_{r} \left(\dfrac{\pi D_r}{\lambda}\right)^2, 
\end{eqnarray}
where $\eta_{bs-t}, \eta_{bs-r}$ and $\eta_{r}$ are the corresponding aperture efficiencies, respectively.

\subsubsection{Path Loss}
According to the free-space path loss (FSPL) model, the path loss of the THz signal between the BS and the user can be described by the FSPL factor as
\begin{equation}
\label{eqn-PL-FF}
PL(d) = \dfrac{\lambda^2}{(4\pi d)^2},\ d \geq \max\left(1, d_R \right),
\end{equation}
where $d$ is the distance between the BS and the user, $\max(\cdot)$ takes the maximum of $1$ and $d_R$, $d_R = 2{D_{bs-t}}^2/\lambda$ is the Rayleigh distance that distinguishes near-field from far-field regions \cite{IEEE-standard-2014, Ahmed2020, LXiao2021, YLiu2023}. Therefore, the far-field operation region can be determined as $d \geq d_R$. The near-field region can be further divided into two regions, i.e., the reactive and radiating near-field regions, also known as the Fresnel zone (FZ) \cite{YLiu2023}, and the boundary to distinguish these two regions is given by $d_r = 0.62\sqrt{{D_{bs-t}}^3/\lambda}$ \cite{Ahmed2020}. As a result, the reactive near-field region and the FZ region can be derived as
\begin{eqnarray}
\label{eqns-Near-Field-Two-Regions}
\begin{cases}
d \in (0, d_r], \ \text{reactive region}, \\
d \in \lbrack d_r, d_R \rbrack \ \text{radiating region (FZ)}.
\end{cases}
\end{eqnarray}
Note that the reactive near-field stores and releases energy rather radiating the energy from the antenna and thus, if $d$ is in this region, the received power at the receiver will be 0. Therefore, we mainly focus on the FZ region and the far-field region.

In the far-field, the path loss of the THz signal has been given in \eqref{eqn-PL-FF}, while in the FZ region, the path loss factor is given in \cite{Ahmed2020} and \cite{IKim2013} as 
\begin{equation}
\label{eqn-PL-NF-FZ}
PL_{FZ}(d) = \dfrac{\lambda^2\gamma_A}{(4\pi d)^2}, \ d \in \lbrack d_r, d_R \rbrack,
\end{equation}
where $\gamma_A \geq 0$ is the reduction factor that accounts for the impact of the FZ operation on the far-field Friss transmission equation in \eqref{eqn-PL-FF}. According to \cite{IKim2013} , the value of $\gamma_A$ is given by
\begin{align}
\label{eqns-Near-Field-Gamma-Values}
\gamma_A &= 
\begin{cases} 
1 - \alpha_E\left(\dfrac{{\pi^2} d}{2\lambda G_{bs-t}}\right)^{-2},\ G_{bs-t} \geq 10\ dB, \\ 
1 - \alpha_E\left(\dfrac{{\pi^2} d}{2\lambda (2G_{bs-t})}\right)^{-2},\ G_{bs-t} < 10\ dB,
\end{cases}
\end{align}
where $\alpha_E = 0.06$ is an empirical gain reduction coefficient for any type of antenna \cite{Ahmed2020, IKim2013}. Since $\gamma_A \geq 0$, the minimum value of $d$ in \eqref{eqns-Near-Field-Gamma-Values} must satisfy
\begin{eqnarray}
\label{eqns-Near-Field-Gamma-Minimum-d}
\begin{cases}
d \geq d_{min} = \dfrac{2 \lambda\sqrt{\alpha_E}G_{bs-t}}{\pi^2}, \ G_{bs-t} \geq 10\ dB, \\
d \geq d_{min} = \dfrac{4 \lambda\sqrt{\alpha_E}G_{bs-t}}{\pi^2}, \ G_{bs-t} < 10\ dB.
\end{cases}
\end{eqnarray}
As a result, $d$ in \eqref{eqn-PL-NF-FZ} can be expressed more rigorously as $d \in \lbrack \max(d_{min}, d_r), d_R \rbrack$. Putting \eqref{eqn-PL-FF} and \eqref{eqn-PL-NF-FZ} together, the path loss factor is
\begin{align}
\label{eqn-PL}
PL(d) &= 
\begin{cases} 
\dfrac{\lambda^2}{(4\pi d)^2},\ d \geq \max\left(1, d_R \right), \\
\dfrac{\lambda^2\gamma_A}{(4\pi d)^2}, \ d \in \lbrack \max(d_{min}, d_r), d_R \rbrack.
\end{cases}
\end{align}

\subsubsection{Atmospheric/Molecular Absorption} 
In addition to path loss, THz signals also attenuate due to molecular absorption \cite{J.M.Jornet2024, W.Jiang2024, J.M.Jornet2011, Kokkoniemi2021, ITU-R2022}. According to \cite{J.M.Jornet2011} and \cite{Kokkoniemi2021}, the molecular absorption loss at a distance of $d$ can be obtained by using the Beer-Lambert Law as \begin{equation}
\label{eqn-MA}
h_{abs} = e^{-k(f)d},\ 100\ GHz \leq f \leq 450\ GHz,
\end{equation}
where $k(f)$ is the medium absorption factor at a given frequency $f$, and it is given in \cite{Kokkoniemi2021} as
\begin{equation}
\label{eqn-MAE}
k(f) = \sum_{i=1}^{6} y_i(f, \mu) + g(f, \mu),
\end{equation}
where $y_i(f, \mu)$ is an absorption coefficient for the $i$-th absorption line, $g(f, \mu)$ is a polynomial to fit the expression to the actual theoretical response, and $\mu$ is the volume mixing ratio of water vapor. More details can be found in \cite{Kokkoniemi2021}.

\subsubsection{Beam Misalignment}
The wavelength of THz signals is much shorter than that of millimeter waves and microwaves, leading to a highly directional propagation. As illustrated in Fig. \ref{fig1-system-design}, it shows a highly directional and narrow beam (known as pencil-like beam) between the BS and the user. To establish a reliable connection between the BS and the user, it is of great importance to address and study beam alignment \cite{W.Jiang2024}, whether for sensing, communications, or powering. According to \cite{B.Chang2022, A.A.Farid2007, A.A.A.Boulogeorgos2019, H.Sarieddeen2021}, the misalignment fading coefficient can be expressed as
\begin{equation}
\label{eqn-MISA}
h_{mis} = S_0e^{-\frac{2{l_{mis}}^2}{{R_{ebw}}^2}},
\end{equation} 
where $l_{mis}$ is the misalignment error expressed as the radial distance between the beam centre of the BS and the beam centre of the receiver at distance $d$, $R_{ebw}$ represents the equivalent beam width, $S_0$ is the fraction of power collected at the receiver when $l_{mis} = 0$ (i.e., complete alignment) with
\begin{equation}
\label{eqn-S_0}
S_0 = \mathrm{erf}(\epsilon)^2,
\end{equation}
$\mathrm{erf}(\cdot)$ is the Gaussian error function, $\epsilon = (\sqrt{\pi}r)/(\sqrt{2}R_d)$, $r$ is the radius of the receiver effective area, and $R_d$ is the beam radius\footnote{Some works also call this the beam waist \cite{A.A.Farid2007}\cite{A.A.A.Boulogeorgos2022} or beam width \cite{B.E.A.Saleh1991}. If not specified otherwise, they are interchangeable in this paper.} at a distance of $d$. According to \cite{B.Chang2022, A.A.Farid2007, A.A.A.Boulogeorgos2019, A.A.A.Boulogeorgos2022}, ${R_{ebw}}^2$ is related to ${R_d}^2$ as 
\begin{equation}
\label{eqn-R_eq}
{R_{ebw}}^2 = {R_d}^2\dfrac{\sqrt{\pi}\mathrm{erf}(\epsilon)}{2\epsilon e^{-\epsilon^2}}.
 \end{equation}
 
\subsubsection{Multi-path Fading}
Since a LoS path is considered between the BS and the receiver, the multi-path fading coefficient can be modeled as a Rician random variable, i.e., $h_f \sim Rice (\nu, \sigma)$ with Rician factor $K = {\nu^2}/{2\sigma^2}$ \cite{A.A.A.Boulogeorgos2022, D.Serghiou2022}, where $\nu$ is the average value or amplitude of the line of sight (LoS) signal, $\sigma$ is the standard deviation of the scattering signals. As a result, the probability density function (PDF) of $h_f$ can be expressed as
\begin{equation}
\label{eqn-h_f_PDF}
f(h_f; \sigma) = \frac{h_f}{\sigma^2} \exp\left(-\frac{{h_f}^2 + 2\sigma^2}{2\sigma^2}\right) I_0\left(\frac{h_f \sqrt{2\sigma^2}}{\sigma^2}\right),
\end{equation}
 where $\quad h_f \geq 0$, and $I_0(\cdot)$ is the modified Bessel function of zero order.
 
\subsubsection{Beam Collection Efficiency}
According to \cite{C.T.Rodenbeck2021}, the beam collection efficiency is defined as the ratio of the effective beam received by the aperture of the receiving antenna to the total beam emitted by the transmitting antenna. Using the Gaussian beam approximation, the beam collection efficiency is given as \cite{C.T.Rodenbeck2021, P.F.Goldsmith1992}
\begin{equation}
\label{eqn-Efficiency-beam}
\eta_b = 1 - e^{-\frac{A_{tx}A_{rx}}{\lambda^2d^2}},
 \end{equation}
where $A_{tx}$ and $A_{rx}$ are transmit and receive aperture areas of the BS and the user, respectively. From \eqref{eqn-Efficiency-beam}, one sees that, when both aperture areas are fixed, the main factors affecting $\eta_b$ are $\lambda$ ($\lambda$ = $c/f$) and $d$. In other words, a higher $f$ leads to a larger $\eta_b$, and a larger $d$ leads to a smaller $\eta_b$. Therefore, $f$ and $d$ need to be chosen carefully in THz-ISCAP.

\subsubsection{RF-to-DC Conversion Loss}
There have been quite a few works on the modeling of the RF to DC conversion efficiency, and both linear and non-linear models were obtained \cite{S.Mizojiri2018, C.Psomas2024, S.Mizojiri2019, N.Shanin2023, Y.Chen2017, R.Jiang2024}. However, the conversion efficiency depends on the input RF power, and it is non-linear in most cases. One has a relationship between the input RF power $x$ and output DC power $f(x)$ as \cite{Y.Chen2017}
\begin{equation}
\label{eqn-Efficiency-RF-2-DC-nonlinear-model}
f_\eta(x) = \dfrac{a_0x+b_0}{x+c_0} - \dfrac{b_0}{c_0},
\end{equation}
where $a_0$, $b_0$ and $c_0$ are constants derived by standard curve fitting. As a result, the non-linear RF-to-DC conversion efficiency at the receiver can be expressed as
\begin{equation}
\label{eqn-Efficiency-RF-2-DC-nonlinear} 
\eta = \dfrac{f_\eta(P_{RF})}{P_{RF}}. 
\end{equation}

\subsection{THz Sensing, Communications and Powering}
\subsubsection{THz Sensing}
It is assumed that the BS sends THz signals to the user with transmit power $P_t$ through a horn antenna. Using \eqref{eqn-PL}, \eqref{eqn-MA}, \eqref{eqn-MISA}, \eqref{eqn-Efficiency-beam} and \eqref{eqn-Efficiency-RF-2-DC-nonlinear}, the signal power received at the user can be expressed as
\begin{equation}
\label{eqn-Received Power}
P_r = P_t G_{bs-t} G_{r} PL(d) h_{abs} |h_{mis}|^2 |h_f|^2 \eta_b.
\end{equation}
For sensing, this signal is reflected by the user and then the reflected signal is received by the BS as
\begin{equation}
\label{eqn-Reflected Power}
P_{r-r}= P_t G_{bs-t} G_{bs-r} PL(2d) {h_{abs}}^2 |h_{mis}|^4 |h_f|^4 {\eta_b}^2.
\end{equation}

\subsubsection{THz Communications} 
Using \eqref{eqn-Received Power}, the channel capacity at the user can be written as
\begin{equation}
\label{eqn-Channel-Capacity}
C = \log_2\left(1+\dfrac{P_r}{N_r}\right),
\end{equation}
where $N_r$ is the noise power at the user.

\subsubsection{THz Powering}
Again, using \eqref{eqn-Efficiency-RF-2-DC-nonlinear-model}, \eqref{eqn-Efficiency-RF-2-DC-nonlinear} and \eqref{eqn-Received Power}, the received DC power at the receiver can be expressed as
\begin{equation}
\label{eqn-Received DC Power}
P_{dc} = \eta P_r = f_\eta(P_r).
\end{equation}

The equations in \eqref{eqn-Channel-Capacity} and \eqref{eqn-Received DC Power} assume that the full received power in \eqref{eqn-Received Power} is used in communications or powering. When SWIPT is applied, only a portion will be used.

%
\section{Problem Formulation and Solutions}
\label{PF and Solutions}
In this section, we will study the performance trade-off among THz sensing, communications and powering. As mentioned before, the proposed THz-ISCAP scheme can be described in two phases, i.e., sensing and SWIPT. Within a fixed time $T$, $\rho_0 T$ ($0\leq \rho_0 \leq 1$) is allocated for sensing and $(1-\rho_0)T$ for the SWIPT, where $\rho_0$ is the time allocation ratio between the two phases. In the second phase, $\rho_1$ is used as the ratio of time splitting or power splitting to balance the performance between communications and powering. In the following, we will first model the sensing process for beam alignment. Then, we will formulate and analyze the performance of SWIPT. Finally, we will evaluate the trade-off among sensing, communications and powering.

\subsection{Phase 1: Sensing}
Beam alignment is important for establishing a reliable connection for communications and powering in the later phase. In this phase, sensing for beam alignment is studied. As indicated in \eqref{eqn-MISA}, the value of $h_{mis}$ is determined by $S_0$, $l_{mis}$ and $R_{ebw}$. Both $S_0$ and $R_{ebw}$ can be calculated from $r$ and $R_d$, and $R_d$ can be obtained by using the Gaussian beam formula \cite[eq. (11.2-8)]{B.E.A.Saleh1991} as 
\begin{equation}
\label{eqn-Gaussian-Beams}
R_d = W(d) = W_0 \sqrt{1+\left(\dfrac{d}{d_0}\right)^2},
\end{equation}
where $W_0$ is the beam waist at distance $d=0$, $d_0 = \pi {W_0}^2/\lambda$ is the Rayleigh range. As a result, $l_{mis}$ is the main factor causing misalignment and hence we mainly focus on reducing the value of $l_{mis}$ through sensing. 

\begin{figure}[!t]
\centering
\includegraphics[width=1\linewidth]{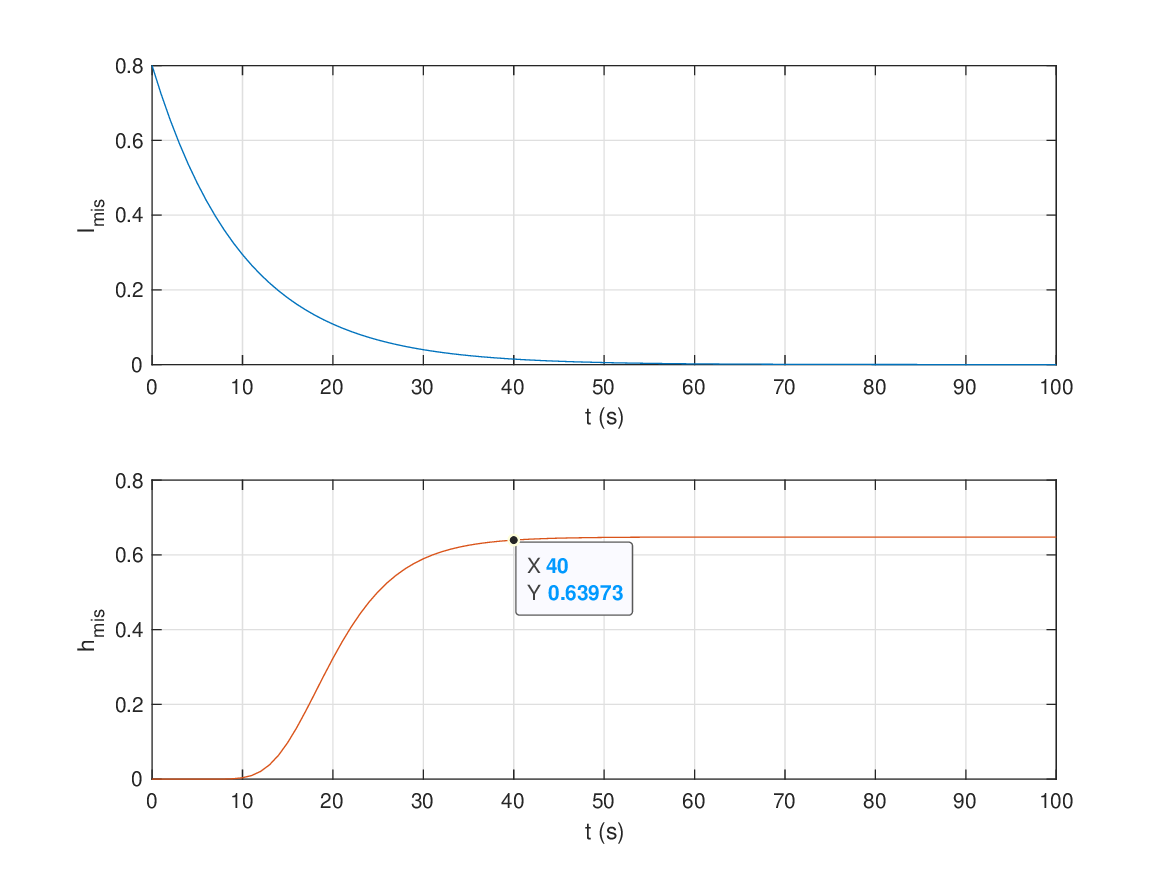}
\caption{Sensing for alignment.}
\label{fig-l-h_mis}
\end{figure}
As in \cite{B.Chang2022}, we assume that the receiver is cellular-connected, and that the receiver can receive control commands from the BS for adjusting its antennas for alignment, i.e., roll, pitch and yaw. At the BS, the reflected signal power $P_{r-r}$ expressed in \eqref{eqn-Reflected Power} is used for sensing. This can be explained as follows. When $P_{r-r}$ is small, it means that the BS should continue sending control commands to the receiver for alignment; when $P_{r-r}$ increases gradually, it suggests $l_{mis}$ is reducing. When $P_{r-r}$ reaches its maximum, it means that $l_{mis}$ reaches its minimum and the optimal alignment is achieved. However, the process of control and adjustment is complex and challenging and thus, some works model the process by using a random variable (RV). For example, in \cite{A.A.A.Boulogeorgos2019}, the misalignment error $l_{mis}$ is modeled as a RV following a Rayleigh distribution. While in \cite{A.A.A.Boulogeorgos2022}, $h_{mis}$ is modeled directly as a RV. To simulate the process of adjusting alignment over time in sensing, we simplify the control process by describing $l_{mis}$ as an exponentially decaying model over time. One has
\begin{equation}
\label{eqn-l_mis-exp-decay-model}
l_{mis}(t) = l_0 e^{(-\alpha * t)},\ t = \rho_0 T,\ 0\leq \rho_0 \leq 1,
\end{equation}
where $l_0$ is the initial misalignment error, $\alpha$ is the decreasing rate, and $\rho_0 T$ is the sensing time. For example, Fig. \ref{fig-l-h_mis} shows the simulation process of sensing for alignment over the time. In the upper part of the figure, one sees that the misalignment error $l_{mis}$ shows an exponential downward trend over time and ultimately reaching a stable state approaching 0. Meanwhile, in the lower part of the figure, one can see that the misalignment fading coefficient $h_{mis}$ increases over time and converges to a stable value after about $40\ s$. The adjustment stops when the misalignment is minimized or when a certain threshold is met. In other words, there exists an optimal time allocation ratio ${\rho_0}^{*}$ that satisfies the alignment requirements for the later SWIPT. In addition, the speed of convergence also depends on the initial misalignment error parameter $l_0$ and the decreasing rate $\alpha$.

\subsection{Phase 2: Communications and Powering}
As $\rho_0T$ has been allocated for sensing in Phase 1, $(1-\rho_0)T$ will be used for SWIPT in Phase 2. In this phase, power splitting is adopted, where $\rho_1$ ($0\leq \rho_1 \leq 1$) and $(1-\rho_1)$ are the ratios of the received power for powering and communications, respectively. Using \eqref{eqn-Received Power}, \eqref{eqn-Channel-Capacity} and \eqref{eqn-Received DC Power}, the received energy in watts$\cdot$s (w$\cdot$s) and achievable rate in bits/Hz (bs/Hz) at the receiver can be obtained as
\begin{eqnarray}
E &=& \eta(1-\rho_0)T\rho_1P_r, \label{eqn-Received-Energy} \\
R &=& (1-\rho_0)T \log_2\left( 1+ \dfrac{(1-\rho_1)P_r}{N_r}\right), \label{eqn-Achievable-Rate}
\end{eqnarray}
where $(1-\rho_0)T$ is the communication/powering time in Phase 2, $\rho_1 P_r$ is the portion of power for energy harvesting and $(1-\rho_1) P_r$ is the portion of power for information delivery.

\subsection{Problems}
From \eqref{eqn-Received-Energy} and \eqref{eqn-Achievable-Rate}, $E$ and $R$ can be seen as a function of $\rho_0$ ($0\leq \rho_0 \leq 1$) and $\rho_1$ ($0\leq \rho_1 \leq 1$). Denote $\mathcal{D} = \{(\rho_0, \rho_1) | \rho_0\in[0, 1], \rho_1\in[0,1]\}$, $E = f_E(\rho_0, \rho_1)$ and $R = f_R(\rho_0, \rho_1)$. The objective is to maximize $f_E(\rho_0, \rho_1)$ and $f_R(\rho_0, \rho_1)$. Even though $\mathcal{D}$ is a closed and finite convex set, $f_E(\rho_0, \rho_1)$ and $f_R(\rho_0, \rho_1)$ may not necessarily be a convex (or concave) function because their Hessian matrices may not be semi-positive (or semi-negative) definite, especially when $\rho_0$ and $\rho_1$ are close to 0. Taking $f_E(\rho_0, \rho_1)$ as an example, ${\partial {f_E}^{2}}/{\partial^2\rho_1} = 0$, but ${\partial {f_E}^{2}}/{\partial\rho_0\rho_1} = {\partial {f_E}^{2}}/{\partial\rho_1\rho_0} \neq 0$. As a result, the eigenvalues of the Hessian matrix corresponding to the function $f_E(\rho_0, \rho_1)$ are not all positive. Next, we will look for their numerical solutions.

\subsubsection{Analysis of $E = f_E(\rho_0, \rho_1)$} For $f_E(\rho_0, \rho_1)$, when fixing $\rho_0$, the first-order partial derivative of $f_E(\rho_0, \rho_1)$ with respect to $\rho_1$ can be derived as
\begin{equation}
\label{eqn-FOPD-f_E_rho1}
\dfrac{\partial f_E}{\partial \rho_1} = (1-\rho_0)\xi_1 e^{\frac{-4{l_0}^{2}e^{-2\alpha\rho_0T}}{R_{ebw}^{2}}},
\end{equation}
where $\xi_1 = \eta T P_t G_{bs-t} G_{r} PL(d) h_{abs} |h_f|^2 \eta_b {S_0}^{2}$ can be seen as a constant greater than 0. As $0\leq \rho_0 \leq 1$, one has ${\partial f_E}/{\partial \rho_1} \geq 0$ (the equality sign only holds when $\rho_0 = 1$, but this is not meaningful in our case), and hence, when $\rho_0 < 1$, $f_E(\rho_0, \rho_1)$ is a monotonically increasing function of $\rho_1$. The maximum value of $f_E(\rho_0, \rho_1)$ will be obtained when $\rho_1$ reaches its maximum. When fixing $\rho_1$, the first-order partial derivative of $f_E(\rho_0, \rho_1)$ with respect to $\rho_0$ can be derived as
\begin{equation}
\label{eqn-FOPD-f_E_rho0}
\dfrac{\partial f_E}{\partial \rho_0} = \rho_1 \xi_1 e^{\frac{-4{l_0}^{2}e^{-2\alpha\rho_0T}}{R_{ebw}^{2}}} \left(\frac{(1-\rho_0)8\alpha T{l_0}^{2}e^{-2\alpha\rho_0T}}{R_{ebw}^{2}}-1\right).
\end{equation}
Let ${\partial f_E}/{\partial \rho_0} = 0$, one has $\rho_1 = 0$, which is ignored as it is a trivial case, and $(1-\rho_0)8\alpha T{l_0}^{2}e^{-2\alpha\rho_0T} = R_{ebw}^{2}$ which is complex to be solved. Denoting $g(\rho_0) = (1-\rho_0)8\alpha T{l_0}^{2}e^{-2\alpha\rho_0T} - R_{ebw}^{2}$, one has
\begin{align}
{\partial g}/{\partial \rho_0} &= - [8\alpha T{l_0}^{2} + (1-\rho_0)16\alpha^2 T^2 {l_0}^{2}]e^{-2\alpha\rho_0 T}, \\
g(0) &= 8\alpha T{l_0}^{2} - R_{ebw}^{2}, \label{eqn-g_rho_0-a} \\
g(1) &= - R_{ebw}^{2} < 0. \label{eqn-g_rho_0-b}
\end{align}
It is found that ${\partial g}/{\partial \rho_0} <0$, and $g(0) > 0$ holds in general and thus, there exists a ${\theta_1}^{\prime} \in (0, 1)$ which makes $g({\theta_1}^{\prime}) = 0$ according to the Zero Theorem. As a result, one can conclude that, when fixing $\rho_1$, $f_E(\rho_0, \rho_1)$ increases with $\rho_0$ first, and then decreases, which means the maximum value of $f_E(\rho_0, \rho_1)$ will be obtained at $\rho_0 = {\theta_1}^{\prime}$. To derive ${\theta_1}^{\prime}$, it can be calculated numerically using the bisection method. According to the above analysis, the maximum value of $f_E(\rho_0, \rho_1)$ on a given subset can be obtained by implementing a divide-and-conquer algorithm. 

\subsubsection{Analysis of $R = f_R(\rho_0, \rho_1)$} Following the same approach, we next examine $f_R(\rho_0, \rho_1)$. When fixing $\rho_0$, one has the first-order partial derivative of $f_R(\rho_0, \rho_1)$ with respect to $\rho_1$ as
\begin{equation}
\label{eqn-FOPD-f_R_rho1}
\dfrac{\partial f_R}{\partial \rho_1} = \dfrac{-\xi_2(1-\rho_0)Te^{\frac{-4{l_0}^{2}e^{-2\alpha\rho_0T}}{R_{ebw}^{2}}}}{\left(1+(1-\rho_1)\xi_2e^{\frac{-4{l_0}^{2}e^{-2\alpha\rho_0T}}{R_{ebw}^{2}}}\right)\ln(2)},
\end{equation}
where $\xi_2 = P_t G_{bs-t} G_{r} PL(d) h_{abs} |h_f|^2 \eta_b {S_0}^{2}/N_r$ can be treated as a positive constant. Let ${\partial f_R}/{\partial \rho_1} =0$, one has $\rho_0 = 1$, which is ignored as discussed before. Hence, one can see that, when $\rho_0 < 1$, ${\partial f_R}/{\partial \rho_1} < 0$, which means $f_R(\rho_0, \rho_1)$ is a monotonically decreasing function of $\rho_1$. Therefore, given $\rho_0$, the maximum value of $f_R(\rho_0, \rho_1)$ can be achieved when $\rho_1$ reaches its minimum. Next, fixing $\rho_1$, one has the first-order derivative of $f_R(\rho_0, \rho_1)$ with respect to $\rho_0$ as
\begin{align}
\label{eqn-FOPD-f_R_rho0}
\dfrac{\partial f_R}{\partial \rho_0} = & -T\log_2\left( 1+ (1-\rho_1)\xi_2 e^{\frac{-4{l_0}^{2}e^{-2\alpha \rho_0 T}}{R_{ebw}^{2}}} \right), \nonumber \\
&+ \dfrac{(1-\rho_0)(1-\rho_1)\xi_2 8\alpha T^2 l_0^2 e^{-2\alpha \rho_0 T} e^{\frac{-4{l_0}^{2}e^{-2\alpha\rho_0T}}{R_{ebw}^{2}}}  }{ R_{ebw}^{2} \left( 1 + (1-\rho_1)\xi_2 e^{\frac{-4{l_0}^{2}e^{-2\alpha\rho_0T}}{R_{ebw}^{2}}}\right)\ln(2) }.
\end{align}
Letting ${\partial f_R}/{\partial \rho_0} = 0$, one has $\rho_1 = 1$, which is ignored as discussed earlier. Denote $h(\rho_0) = {\partial f_R}/{\partial \rho_0}$. One has
\begin{align}
\label{eqn-h_rho_0}
&{\partial h}/{\partial \rho_0} = - \left(1+(1-\rho_1)\xi_2 e^{\frac{-4{l_0}^{2}e^{-2\alpha \rho_0 T}}{R_{ebw}^{2}}} \right)\dfrac{8\alpha T^2{l_0}^{2}}{\ln(2)R_{ebw}^{2}} \nonumber \\
& \qquad \qquad \qquad \times (1-\rho_1)\xi_2 e^{\frac{-4{l_0}^{2}e^{-2\alpha \rho_0 T}}{R_{ebw}^{2}}} e^{-2\alpha \rho_0 T} \nonumber \\
& \qquad \qquad \ - \dfrac{8\alpha T^2{l_0}^{2}  [1+(1-\rho_0)2\alpha T] }{\left(1+ (1-\rho_1)\xi_2 e^{\frac{-4{l_0}^{2}e^{-2\alpha \rho_0 T}}{R_{ebw}^{2}}}\right)\ln(2)R_{ebw}^{2} } \nonumber \\
& \qquad \qquad \qquad \times (1-\rho_1)\xi_2 e^{\frac{-4{l_0}^{2}e^{-2\alpha \rho_0 T}}{R_{ebw}^{2}}} e^{-2\alpha \rho_0 T} \nonumber \\
& \qquad \qquad \ + \dfrac{(1-\rho_0)(1-\rho_1)\xi_2 64\alpha^2 T^3 {l_0}^{4}}{\left(1+ (1-\rho_1)\xi_2 e^{\frac{-4{l_0}^{2}e^{-2\alpha \rho_0 T}}{R_{ebw}^{2}}}\right)^2 \ln(2) R_{ebw}^{4} },\\
&h(0)=\dfrac{(1-\rho_1)\xi_2 8\alpha T^2 l_0^2 e^{\frac{-4{l_0}^{2}}{R_{ebw}^{2}}} }{ R_{ebw}^{2} \left( 1 + (1-\rho_1)\xi_2 e^{\frac{-4{l_0}^{2}}{R_{ebw}^{2}}}\right)\ln(2) }  \nonumber\\
&\qquad\quad -T\log_2 \left(1+ (1-\rho_1)\xi_2 e^{\frac{-4{l_0}^{2}e^{-2\alpha \rho_0 T}}{R_{ebw}^{2}}} \right), \\
&h(1) = -T\log_2 \left( 1+ (1-\rho_1)\xi_2 e^{\frac{-4{l_0}^{2}e^{-2\alpha T}}{R_{ebw}^{2}}} \right)< 0.
\end{align}
It is found that ${\partial h}/{\partial \rho_0} < 0$ and $h(0) > 0$ holds in general and thus, similar to $f_E(\rho_0, \rho_1)$ above (see \eqref{eqn-g_rho_0-a} and \eqref{eqn-g_rho_0-b}), one can conclude that, when fixing $\rho_1$, $f_R(\rho_0, \rho_1)$ increases with $\rho_0$ and then decreases. Therefore, there exists a ${\theta_2}^{\prime}\in (0, 1)$ which makes $h({\theta_2}^{\prime})=0$, and the maximum value of $f_R(\rho_0, \rho_1)$ can be achieved at $\rho_0 = {\theta_2}^{\prime}$ when fixing $\rho_1$. The bisection method can be used to derive ${\theta_2}^{\prime}$ numerically, and the maximum value of $f_R(\rho_0, \rho_1)$ on a given subset can be obtained by designing a divide-and-conquer algorithm.

However, note that, when fixing $\rho_0$, $f_E$ and $f_R$ are monotonically increasing and decreasing functions of $\rho_1$, respectively, which means their maximum values cannot be obtained simultaneously. As a result, we maximize one with a minimum requirement on the other, i.e., maximize $f_E(\rho_0, \rho_1)$ with $f_R(\rho_0, \rho_1) \geq R_{\epsilon}$ ($0 \leq R_{\epsilon}\leq \max(f_R)$) or maximize $f_R(\rho_0, \rho_1)$ with $f_E(\rho_0, \rho_1) \geq E_{\epsilon}$ ($0 \leq E_{\epsilon}\leq \max(f_E)$).

\begin{algorithm}[t]
\caption{Maximization of \(E = f_E(\rho_0, \rho_1)\)}
\begin{algorithmic}[1]
\label{alg-P1}
\STATE \textbf{Input:} Convex function \(f_E(\rho_0, \rho_1)\), $R_\epsilon$
\STATE \textbf{Output:} Optimal value \(E^* = f({\rho_0}^*, {\rho_1}^*)\), and \(({\rho_0}^*, {\rho_1}^*)\) \\
\textcolor{blue}{// Calculate convex set \(S_1 \) using \eqref{eqn-RgeqRepsilon2-2}}
\WHILE{Stopping criterion not met}
    \STATE Find the extreme point of the function in \eqref{eqn-RgeqRepsilon2-2}
    \STATE $\mu^* \gets {(2^{\frac{R_{\epsilon}}{(1-\rho_0)T}}-1)}/{C_1e^{-\frac{4{l_0}^{2}e^{-2 \alpha \rho_0T}}{{R_{ebw}}^2}}} $
\ENDWHILE
\WHILE{Stopping criterion not met}
    \STATE Find $p_1 \in [0, \mu^*]$
    \STATE $p_1 \gets {(2^{\frac{R_{\epsilon}}{(1-\rho_0)T}}-1)}/{C_1e^{-\frac{4{l_0}^{2}e^{-2 \alpha \rho_0T}}{{R_{ebw}}^2}}} = 1 $
\ENDWHILE
\WHILE{Stopping criterion not met}
    \STATE Find $p_2 \in [\mu^*, 1]$
    \STATE $ p_2 \gets {(2^{\frac{R_{\epsilon}}{(1-\rho_0)T}}-1)}/{C_1e^{-\frac{4{l_0}^{2}e^{-2 \alpha \rho_0T}}{{R_{ebw}}^2}}} = 1 $
\ENDWHILE \\
\textcolor{blue}{// Calculate the maximum $f_E(\rho_0, \rho_1)$}
\STATE Calculate convex set \(\mathcal{D}_1 \subseteq \mathbb{R}^2\) using \eqref{eqn-RgeqRepsilon-subset}
\FOR {${\rho_0}^{\prime}$ in $[0,1) \cap S_1$}
\STATE ${\rho_1}_{max} \gets {\rho_0}^{\prime}$ using the inequality in \eqref{eqn-RgeqRepsilon-subset}
\STATE Calculate $f_E({\rho_0}^{\prime}, {\rho_1}_{max})$ and update with the max
\ENDFOR
\STATE ${\rho_1}^*, {\rho_0}^* \gets \max {f_E({\rho_0}^{\prime}, {\rho_1}_{max})}$
\STATE Return \(E^* = f_E({\rho_0}^*, {\rho_1}^*)\) and \( ({\rho_1}^*, {\rho_0}^*)\)
\end{algorithmic}
\end{algorithm}
 
\subsubsection{Maximize E with $R \geq R_{\epsilon}$} Let $f_R(\rho_0, \rho_1) \geq R_{\epsilon}$ and using \eqref{eqn-Achievable-Rate}, one has
\begin{equation}
\label{eqn-RgeqRepsilon1}
(1-\rho_0)T \log_2\left( 1+ \dfrac{(1-\rho_1)P_r}{N_r}\right) \geq R_{\epsilon}.
\end{equation}
Then, using \eqref{eqn-RgeqRepsilon1} and $\rho_1 \in [0, 1]$, we have
\begin{eqnarray}
\rho_1 \leq 1 - \dfrac{(2^{\frac{R_{\epsilon}}{(1-\rho_0)T}}-1)}{C_1e^{-\frac{4{l_0}^{2}e^{-2 \alpha \rho_0T}}{{R_{ebw}}^2}}},\ \rho_0 \neq 1, \label{eqn-RgeqRepsilon2-1}\\
\dfrac{(2^{\frac{R_{\epsilon}}{(1-\rho_0)T}}-1)}{C_1e^{-\frac{4{l_0}^{2}e^{-2 \alpha \rho_0T}}{{R_{ebw}}^2}}} \leq 1,\ \rho_0 \neq 1, \label{eqn-RgeqRepsilon2-2}
\end{eqnarray}
where $C_1 = P_t G_{bs-t} G_{r} PL(d) h_{abs} |h_f|^2 \eta_b {S_0}^{2}/ N_r$ can be regarded as a constant. It is challenging to derive a close-form set of $\rho_0$ that meets \eqref{eqn-RgeqRepsilon2-2}, but there exists a closed convex subset of $\rho_0$ because $f_R(\rho_0, \rho_1)$ increases and then decreases with respect to $\rho_0$ when fixing $\rho_1$. Denote $S_1 =\{\rho_0 | \rho_0\in [p_1, p_2] \}$ that satisfies \eqref{eqn-RgeqRepsilon2-2}. One can obtain a new subset of $\mathcal{D}$, which meets $R \geq R_{\epsilon}$, as
\begin{align}
\label{eqn-RgeqRepsilon-subset}
\mathcal{D}_1  = \Bigg\{(\rho_0, \rho_1) |& \rho_0\in[0, 1)\cap S_1, \nonumber \\
& 0 \leq \rho_1\leq1 - \frac{(2^{\frac{R_{\epsilon}}{(1-\rho_0)T}}-1)}{C_1e^{-\frac{4{l_0}^{2}e^{-2 \alpha \rho_0T}}{{R_{ebw}}^2}}} \Bigg\}.
\end{align}
Actually, $[0, 1)\cap S_1 = S_1$ because $0\leq q_1$ and $q_2 <1$. As a result, the optimization problem becomes
\begin{align}
\label{eqn-P1}
\text{(P1):} \quad & \max_{{\rho_0}^{*}, {\rho_1}^{*}}f_E(\rho_0, \rho_1), \\
\text{Subject to:}\quad & \mathcal{D}_1\ \text{in}\ \eqref{eqn-RgeqRepsilon-subset}.
\end{align}
To solve (P1) above, a divide-and-conquer method is used, and details are summarized in Algorithm \ref{alg-P1}.

In Algorithm \ref{alg-P1}, the complexity mainly comes from calculating $S_1$ and finding the maximum $f_E$. For an accuracy of $\varepsilon = 0.01$ in a binary search for $S_1$ and one-dimensional traversal search for $f_E(\rho_0, \rho_1)$, the total complexity can be derived as
\begin{equation}
\label{eqn-Alg1-Complexity}
O (\log_2(\frac{1}{\varepsilon})) + O (\log_2(\frac{\mu^*}{\varepsilon})) + O (\log_2(\frac{1-\mu^*}{\varepsilon})) +O (\frac{p_2 - p_1}{\varepsilon}).
\end{equation}
The maximum $f_E(\rho_0, \rho_1)$ can also be achieved by selecting a ${\rho_1}^{\prime}$ from $\mathcal{D}_1$, and then finding the optimal ${\rho_0}^*$ from $S_1$ which maximizes $f_E(\rho_0, {\rho_1}^{\prime})$. The optimal ${\rho_0}^*$ can be used to calculate the maximum ${\rho_1}_{max}$ using the inequality in \eqref{eqn-RgeqRepsilon-subset}, i.e., ${\rho_1}^* = {\rho_1}_{max}$. Compared to the two-dimensional search whose complexity is $O(\frac{p_2 - p_1}{\varepsilon} \times \frac{\max(\rho_1)-\min(\rho_1)}{\varepsilon})$, the above is much more efficient.

\begin{algorithm}[t]
\caption{Maximization of \(R = f_R(\rho_0, \rho_1)\)}
\begin{algorithmic}[1]
\label{alg-P2}
\STATE \textbf{Input:} Convex function \(f_R(\rho_0, \rho_1)\), $E_\epsilon$
\STATE \textbf{Output:} Optimal value \(R^{*} = f({\rho_0}^{*}, {\rho_1}^{*})\), and \(({\rho_0}^{*}, {\rho_1}^{*})\) \\
\textcolor{blue}{// Calculate convex set $S_{2-1}$ or $S_{2-2}$ using \eqref{eqn-EgeqEepsilon2-2} or \eqref{eqn-EgeqEepsilon3-2}}
\WHILE{Stopping criterion not met}
    \STATE Find the extreme point of the function in \eqref{eqn-EgeqEepsilon2-2} or \eqref{eqn-EgeqEepsilon3-2}
    \STATE ${\nu_1}^{*} \gets \eqref{eqn-EgeqEepsilon2-2}$ or ${\nu_2}^{*} \gets \eqref{eqn-EgeqEepsilon3-2}$
\ENDWHILE
\WHILE{Stopping criterion not met}
    \STATE Find $q_1 \in [0, {\nu_1}^{*}]$ or Find $q_3 \in [0, {\nu_2}^{*}]$
    \STATE $q_1 \gets  \eqref{eqn-EgeqEepsilon2-2}$ or $q_3 \gets  \eqref{eqn-EgeqEepsilon3-2}$
\ENDWHILE
\WHILE{Stopping criterion not met}
    \STATE Find $q_2 \in [{\nu_1}^{*}, 1]$ or Find $q_4 \in [{\nu_2}^{*}, 1]$
    \STATE $ q_2 \gets \eqref{eqn-EgeqEepsilon2-2}$ or $q_4 \gets  \eqref{eqn-EgeqEepsilon3-2}$
\ENDWHILE \\
\textcolor{blue}{// Calculate the maximum $f_R(\rho_0, \rho_1)$}
\STATE Calculate convex set $\mathcal{D}_{2-1}$ (or $\mathcal{D}_{2-2}$) $\subseteq \mathbb{R}^2$ using \eqref{eqn-RgeqRepsilon-subset}
\FOR {${\rho_0}^{\prime}$ in $S_{2-1}$ (or ${\rho_0}^{\prime}$ in $S_{2-2}$)} 
\STATE ${\rho_1}_{min} \gets {\rho_0}^{\prime}$ using the inequality in \eqref{eqn-EgeqEepsilon-subset1} or \eqref{eqn-EgeqEepsilon-subset2}
\STATE Calculate $f_R({\rho_0}^{\prime}, {\rho_1}_{min})$ and update with the max
\ENDFOR
\STATE ${\rho_1}^*, {\rho_0}^* \gets \max {f_R({\rho_0}^{\prime}, {\rho_1}_{min})}$
\STATE Return \(R^{*} = f_R({\rho_0}^{*}, {\rho_1}^{*})\) and \( ({\rho_1}^{*}, {\rho_0}^{*})\)
\end{algorithmic}
\end{algorithm}
\subsubsection{Maximize R with $E \geq E_{\epsilon}$} Let $f_E(\rho_0, \rho_1) \geq E_{\epsilon}$. Using \eqref{eqn-Received-Energy}, one has
\begin{equation}
\label{eqn-EgeqEepsilon1}
 \eta(1-\rho_0)T\rho_1P_r \geq E_{\epsilon}.
\end{equation}
Note that $\eta$ in \eqref{eqn-EgeqEepsilon1} can be linear or non-linear in different scenarios. Next, both cases will be discussed.

\paragraph{$\eta$ is non-linear} Using \eqref{eqn-Efficiency-RF-2-DC-nonlinear-model} and \eqref{eqn-Received DC Power}, \eqref{eqn-EgeqEepsilon1} can be further rewritten as
\begin{equation}
\label{eqn-EgeqEepsilon1-2}
 (1-\rho_0)T\rho_1 \left( \dfrac{a_0 C_2e^{-\frac{4{l_0}^{2}e^{-2 \alpha \rho_0T}}{{R_{ebw}}^2}} +b_0}{C_2e^{-\frac{4{l_0}^{2}e^{-2 \alpha \rho_0T}}{{R_{ebw}}^2}}+c_0} - \dfrac{b_0}{c_0}\right) \geq E_{\epsilon},
\end{equation}
where $C_2 = P_t G_{bs-t} G_{r} PL(d) h_{abs} |h_f|^2 \eta_b {S_0}^{2}$ can be seen as a constant. Using $\rho_1 \in [0, 1]$, \eqref{eqn-EgeqEepsilon1-2} can be simplified by expressing $\rho_1$ in terms of $\rho_0$. One derives
\begin{eqnarray}
\rho_1 \geq \dfrac{E_{\epsilon}/(1-\rho_0)}{ T \left( \dfrac{a_0 C_2e^{-\frac{4{l_0}^{2}e^{-2 \alpha \rho_0T}}{{R_{ebw}}^2}} +b_0}{ C_2 e^{-\frac{4{l_0}^{2}e^{-2 \alpha \rho_0T}}{{R_{ebw}}^2}}+c_0} - \dfrac{b_0}{c_0}\right) },\ \rho_0 \neq 1, \label{eqn-EgeqEepsilon2-1}\\
(1-\rho_0) \left( \dfrac{a_0 C_2e^{-\frac{4{l_0}^{2}e^{-2 \alpha \rho_0T}}{{R_{ebw}}^2}} +b_0}{C_2e^{-\frac{4{l_0}^{2}e^{-2 \alpha \rho_0T}}{{R_{ebw}}^2}}+c_0} - \dfrac{b_0}{c_0}\right) \geq \dfrac{E_{\epsilon}}{ T}. \label{eqn-EgeqEepsilon2-2}
\end{eqnarray}
Denote $S_{2-1} = \{\rho_0 | \rho_0\in [q_1, q_2] \}$ that satisfies \eqref{eqn-EgeqEepsilon2-2}. One can derive a new subset of $\mathcal{D}$ which meets $E \geq E_{\epsilon}$ as
\begin{align}
\label{eqn-EgeqEepsilon-subset1}
&\mathcal{D}_{2-1} = \Bigg\{(\rho_0, \rho_1) | \rho_0\in[0, 1)\cap S_{2-1}, \nonumber \\
&\dfrac{E_{\epsilon}}{ (1-\rho_0)T \left( \dfrac{a_0 C_2e^{-\frac{4{l_0}^{2}e^{-2 \alpha \rho_0T}}{{R_{ebw}}^2}} +b_0}{ C_2 e^{-\frac{4{l_0}^{2}e^{-2 \alpha \rho_0T}}{{R_{ebw}}^2}}+c_0} - \dfrac{b_0}{c_0}\right) } \leq \rho_1 \leq 1 \Bigg\}.
\end{align}
Therefore, the optimization problem can be written as
\begin{align}
\label{eqn-P2-1}
\text{(P2-1):} \quad & \max_{{\rho_0}^{*}, {\rho_1}^{*}}f_R(\rho_0, \rho_1), \\
\text{Subject to:}\quad & \mathcal{D}_{2-1}\ \text{in}\ \eqref{eqn-EgeqEepsilon-subset1}.
\end{align}

\paragraph{$\eta$ is linear} When $\eta$ works in the linear regime, using $\rho_1 \in [0, 1]$, \eqref{eqn-EgeqEepsilon1} can be simplified by expressing $\rho_1$ in terms of $\rho_0$. One derives
\begin{eqnarray}
\rho_1 \geq \dfrac{E_{\epsilon}}{ (1-\rho_0) T C_3e^{-\frac{4{l_0}^{2}e^{-2 \alpha \rho_0T}}{{R_{ebw}}^2}} },\ \rho_0 \neq 1, \label{eqn-EgeqEepsilon3-1}\\
(1-\rho_0) e^{-\frac{4{l_0}^{2}e^{-2 \alpha \rho_0T}}{{R_{ebw}}^2}} \geq \dfrac{E_{\epsilon}}{T C_3}, \label{eqn-EgeqEepsilon3-2}
\end{eqnarray}
where $C_3 = \eta P_t G_{bs-t} G_{r} PL(d) h_{abs} |h_f|^2 \eta_b {S_0}^{2}$ can be seen as a constant. Similarly, denote $S_{2-2} = \{\rho_0 | \rho_0\in [q_3, q_4] \}$ that satisfies \eqref{eqn-EgeqEepsilon3-2}. One derives a new subset of $\mathcal{D}$, which meets $E \geq E_{\epsilon}$ as
\begin{align}
\label{eqn-EgeqEepsilon-subset2}
\mathcal{D}_{2-2} = \Bigg\{&(\rho_0, \rho_1) | \rho_0\in[0, 1)\cap S_{2-2}, \nonumber \\
& \dfrac{E_{\epsilon}}{ T (1-\rho_0) C_3e^{-\frac{4{l_0}^{2}e^{-2 \alpha \rho_0T}}{{R_{ebw}}^2}} } \leq \rho_1 \leq 1 \Bigg\},
\end{align}
and the optimization in this case can be expressed as
\begin{align}
\label{eqn-P2-2}
\text{(P2-2):} \quad & \max_{{\rho_0}^{*}, {\rho_1}^{*}}f_R(\rho_0, \rho_1), \\
\text{Subject to:}\quad & \mathcal{D}_{2-2}\ \text{in}\ \eqref{eqn-EgeqEepsilon-subset2}.
\end{align}
To solve (P2-1) or (P2-2), the same method, including complexity analysis, as solving (P1) can be used. Details are summarized in Algorithm \ref{alg-P2}.

%
%

\begin{table}[t]
\noindent
\begin{tabular}{l | l | l}
\hline
\textbf{Notation} & \textbf{Meaning} & \textbf{Value} \\
\hline
 $f$ & Operating frequency & 300 $GHz$ \\
\hline
 $P_t$ & Transmit power & 10 $W$ \\
\hline
 $D_{bs-t}$ & Aperture diameter of the TX antenna & 0.1 $m$ \\
 \hline
 $D_r$ & Aperture diameter of the RX antenna & 0.2 $m$ \\
 \hline
 $\eta_{bs-t}$ & Aperture efficiency of the TX antenna & 0.2 \\
 \hline
 $\eta_{r}$ & Aperture efficiency of RX antenna & 0.2 \\
 \hline
 $ d $ & The distance between the Tx and the Rx & 20 $m$ \\
 \hline
 $k(f)$ & Medium absorption factor in \eqref{eqn-MA} or \eqref{eqn-MAE} & Same as in \cite{Kokkoniemi2021} \\
 \hline
 $a_0$ & Parameter in \eqref{eqn-Efficiency-RF-2-DC-nonlinear-model} & 0.3929 \cite{Y.Chen2017} \\ 
 \hline
 $b_0$ & Parameter in \eqref{eqn-Efficiency-RF-2-DC-nonlinear-model} & 0.01675 \cite{Y.Chen2017} \\
 \hline
 $c_0$ & Parameter in \eqref{eqn-Efficiency-RF-2-DC-nonlinear-model} & 0.04401 \cite{Y.Chen2017} \\
 \hline
 $l_0 $ & Initial misalignment error & 0.8 \\
 \hline
 $\alpha$ & Decreasing rate & 0.1 \\
 \hline
 $K$ & Rician factor & 1 \\
 \hline
 $T$ & Time for ISCAP & 100 $s$ \\
 \hline
 $N_r$ & Noise power & -50 $dBm$ \\
\hline
\end{tabular}
\vspace{0.5 em}
\caption{Parameter settings for the simulation}
\label{TAB-Parameters}
\end{table}

\section{Numerical Results and Discussion}
\label{Numerical Results}
In this section, simulation results are presented to show the performance trade-off between THz sensing, communications and powering. The parameters used in the simulation are summarized in Table \ref{TAB-Parameters}, if not stated otherwise. For $k(f)$ in the table, the relative humidity and temperature in $k(f)$ are set to $50\%$ and $25^\circ\mathrm{C}$ \cite{Kokkoniemi2021}. The multi-path fading coefficient is generated by using 1000 random values of $|h_f|^2$.

\begin{figure}[t]
\centering
\includegraphics[width=1\linewidth]{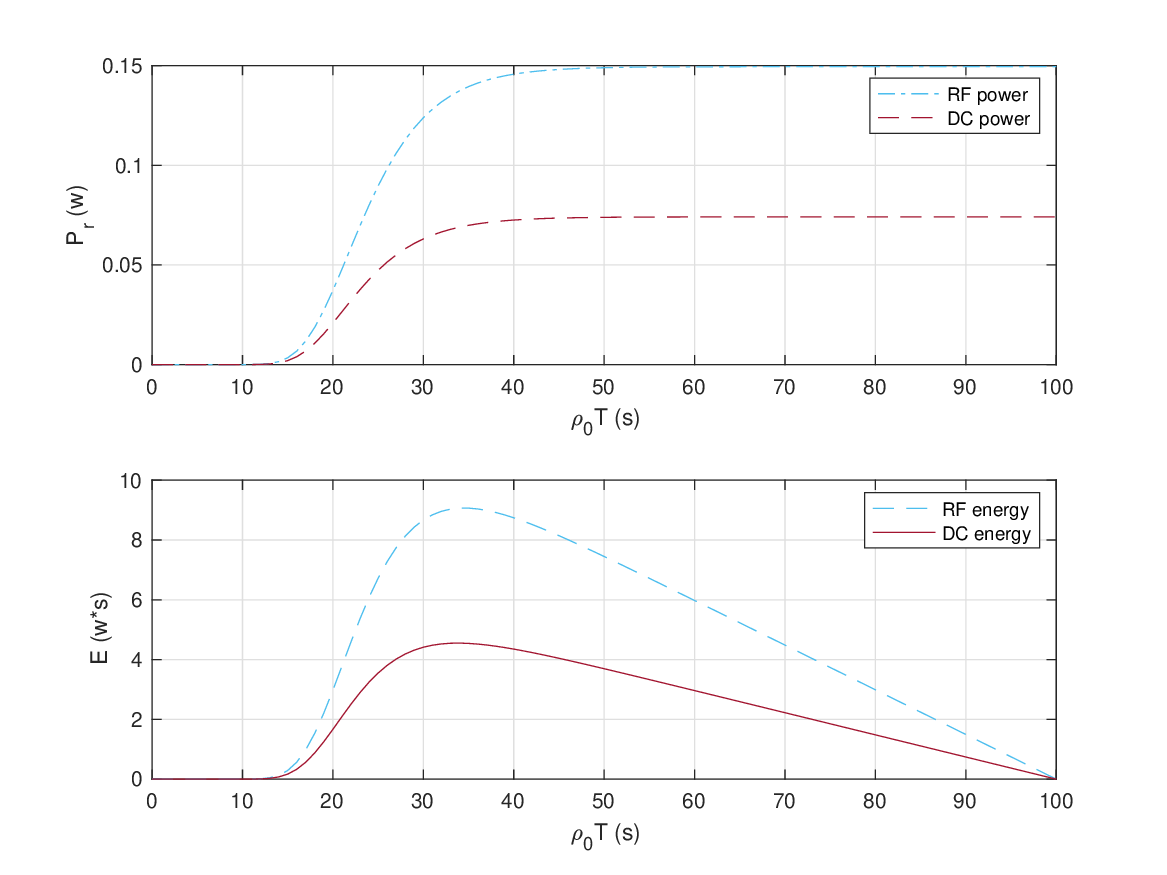}
\caption{$P_r$ and $E$ versus $\rho_0T$.}
\label{fig-RF_DC_PandE}
\end{figure}
Fig. \ref{fig-RF_DC_PandE} shows the received power $P_r$ and the total harvested energy $E$ at the receiver versus the sensing time. In the upper part of the figure, $P_r$ starts at a very small value and then increases, and finally reaches a stable state when $\rho_0 T$ increases. This can be explained as follows. In the beginning, since the misalignment is large,  it causes a significant loss, and the received power is small. Given more sensing time, the misalignment becomes smaller and smaller, leading to an increasing $P_r$. However, when the misalignment error is below a threshold, the received power changes little. In the lower part of the figure, it shows the same trend as the received RF power in the first 33 $s$, but it decreases to 0. This is because the harvested energy increases with the received RF power. However, when the sensing time reaches 100 $s$, there is no time left for harvesting. This is why the received energy becomes 0 in the end. As a result, the balance between sensing in the first phase and communicating \& powering in the second phase is very important. Next, we will investigate the effect of sensing on communications and powering.

\begin{figure}[t]
\centering
\includegraphics[width=1\linewidth]{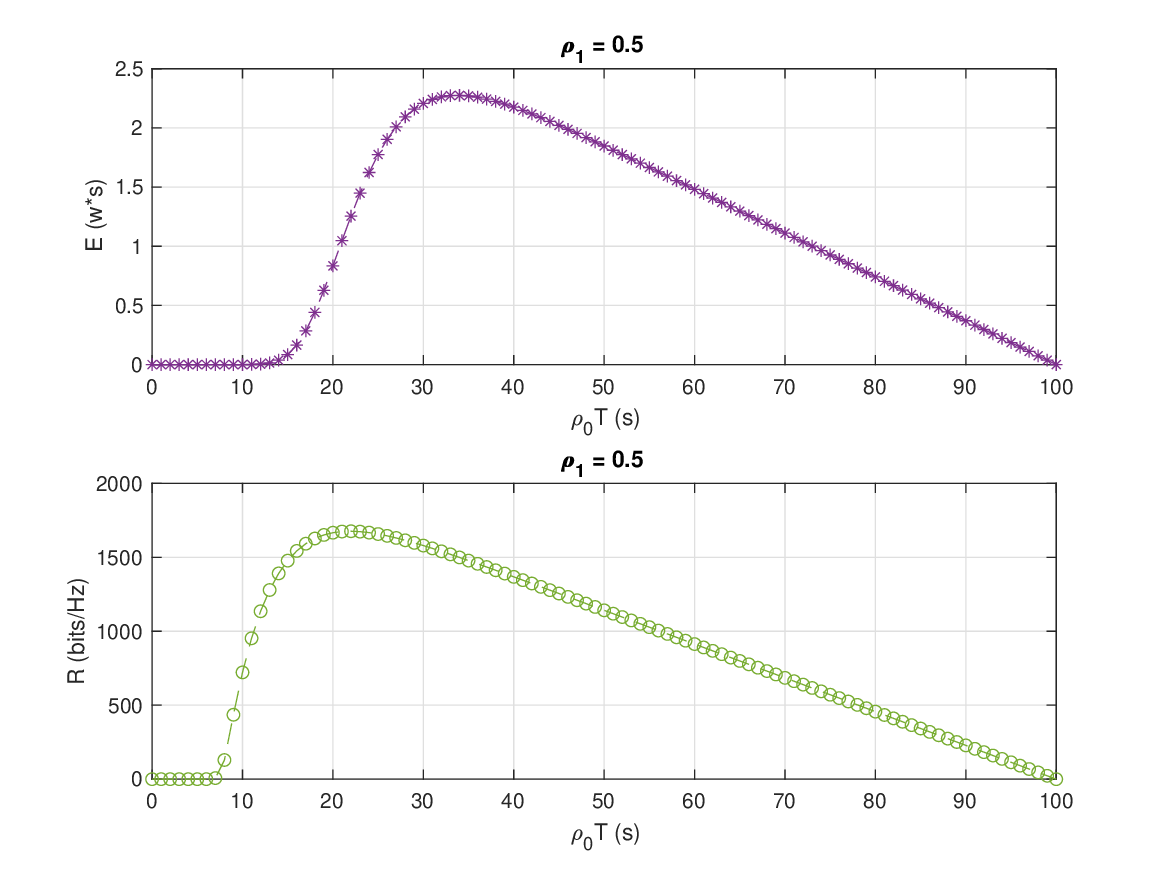}
\caption{$E$ and $R$ vs $\rho_0T$ with $\rho_1 = 0.5$.}
\label{fig-Energy_Communications}
\end{figure}
Fig. \ref{fig-Energy_Communications} shows the harvested energy $E$ and the achievable rate $R$ versus the sensing time $\rho_0 T$. In this figure, $\rho_1$ is set to 0.5. From the figure, both $E$ and $R$ change in the same way as $\rho_0 T$ increases. However, the effect of sensing time $\rho_0 T$ on $E$ and $R$ are slightly different. When the sensing time is set to about 33 $s$, $E$ achieves its maximum in the upper part of the figure, while $R$ reaches its maximum at 21$s$ in the lower part of the figure. Although it is possible to make both reach their maximum simultaneously by adjusting the value of $\rho_1$, this changes their maximum value, or requires reallocation of power using $\rho_1$. Therefore, there is also a balance between communications and powering.

\begin{figure}[t]
   \centering
    \begin{minipage}{0.45\textwidth}
        \centering
        \includegraphics[width=1\linewidth]{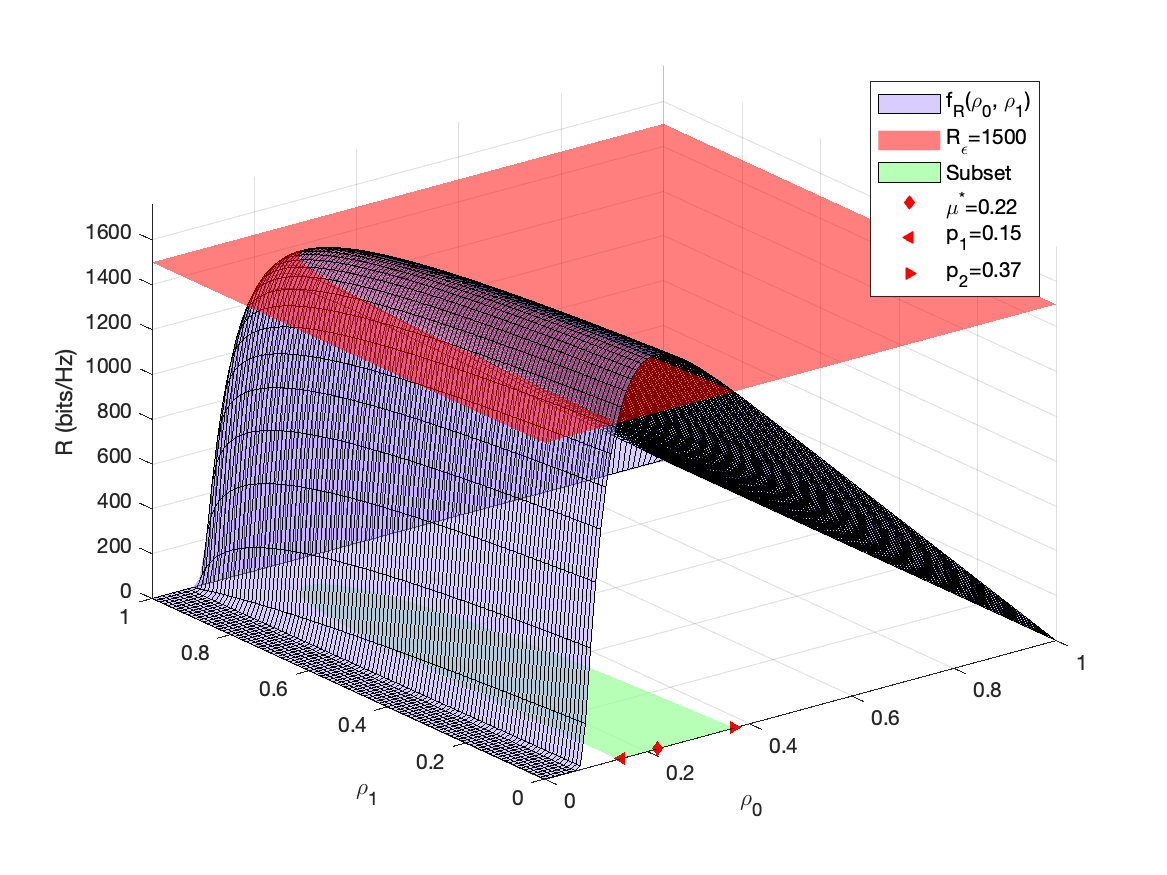}
        \caption*{(a) $R \geq R_{\epsilon} = 1500 \ bits/Hz$.} 
    \end{minipage}
    \hspace{0.5cm} 
    \begin{minipage}{0.45\textwidth}
        \centering
        \includegraphics[width=1\linewidth]{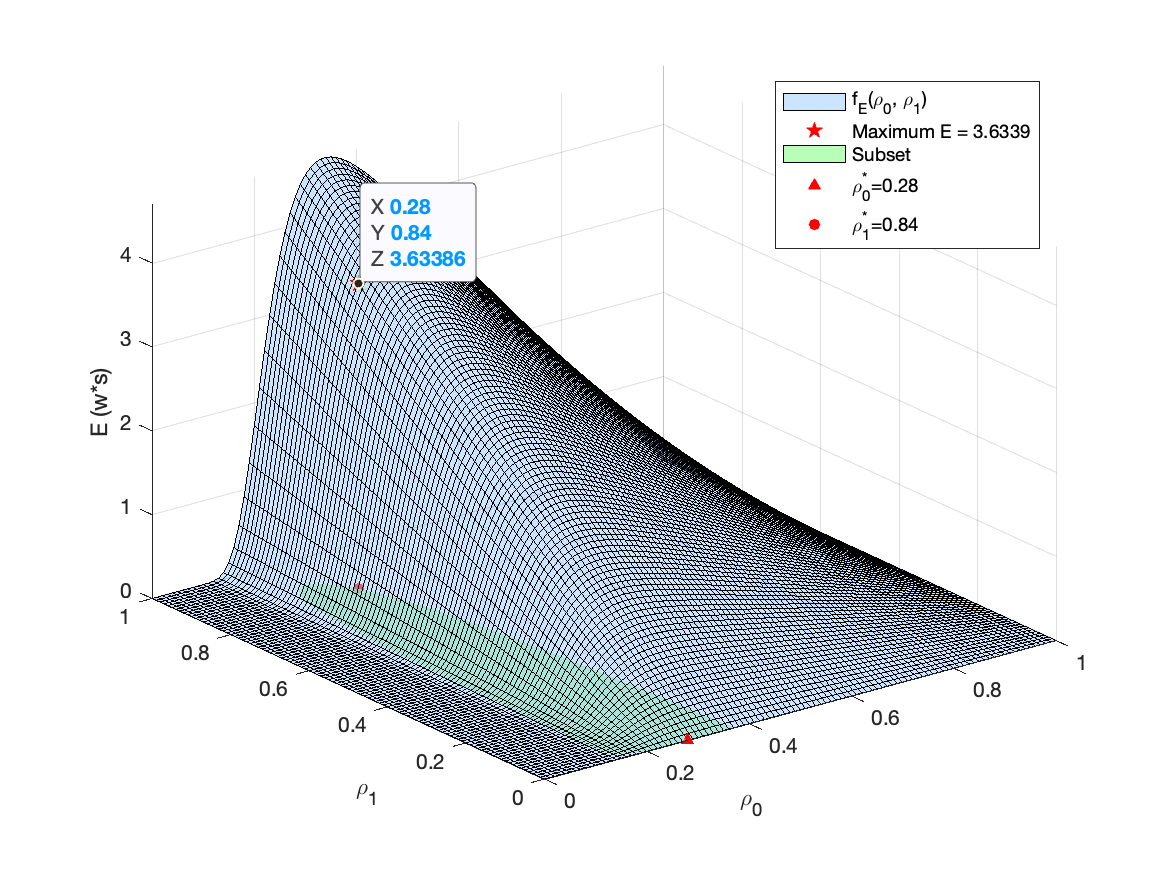} 
        \caption*{(b) Maximum value of $E$.} 
    \end{minipage}
    \caption{Maximize $E$ with $R \geq R_{\epsilon}$.}
    \label{fig-rho_01_MaxEfixC}
\end{figure}
Fig. \ref{fig-rho_01_MaxEfixC} shows the maximization of $E$ with respect to $\rho_0$ and $\rho_1$ for $R \geq R_{\epsilon}$. In this figure, $R_{\epsilon} = 1500\ bits/Hz$ and the step size for binary search is set to 0.01. In Fig. \ref{fig-rho_01_MaxEfixC} (a), one shows that $R$ changes with different $\rho_0$ and $\rho_1$. Particularly, given a $\rho_0$, $R$ decreases with $\rho_1$. This is because $(1-\rho_1)P_r$ is used for communications. Besides, the plane shows the locations where $R_{\epsilon} = 1500\ bits/Hz$, by which a subset of $\mathcal{D}$ that satisfies $R \geq R_{\epsilon}$ is derived shown as the shadow area in the $\rho_0\textendash\rho_1$ plane. According to Algorithm \ref{alg-P1}, the values of $\mu^*$, $p_1$ and $p_2$ are obtained as $0.22$, $0.15$ and $0.37$, respectively, as indicated by the diamond, the left triangle, and the right triangle in the figure too. Then, Fig. \ref{fig-rho_01_MaxEfixC} (b) shows the maximum value of $E$ for this subset. One can see that its maximum $E^{*}$ is 3.63386 $(w*s)$, and the optimal sensing time ratio ${\rho_0}^*$ and power splitting ratio ${\rho_1}^*$ are $0.28$ and $0.84$, respectively, as indicated by the upper triangle and the solid circle in this case. Note that the maximum value of $E$ is not necessarily obtained at the top point of the subset area. In Fig. \ref{fig-rho_01_MaxEfixC} (a), $\mu^{*}=0.22$ marked the value of $\rho_0$ that can make $\rho_1$ take its maximum in the subset, but this does not mean that $E$ can take the maximum at the same point. Instead, $E$ reaches its maximum when $\rho_0^{*}=0.28$ which is slightly greater than $\mu^{*}=0.22$.

\begin{figure}[t]
    \centering
    \begin{minipage}{0.45\textwidth}
        \centering
        \includegraphics[width=1\linewidth]{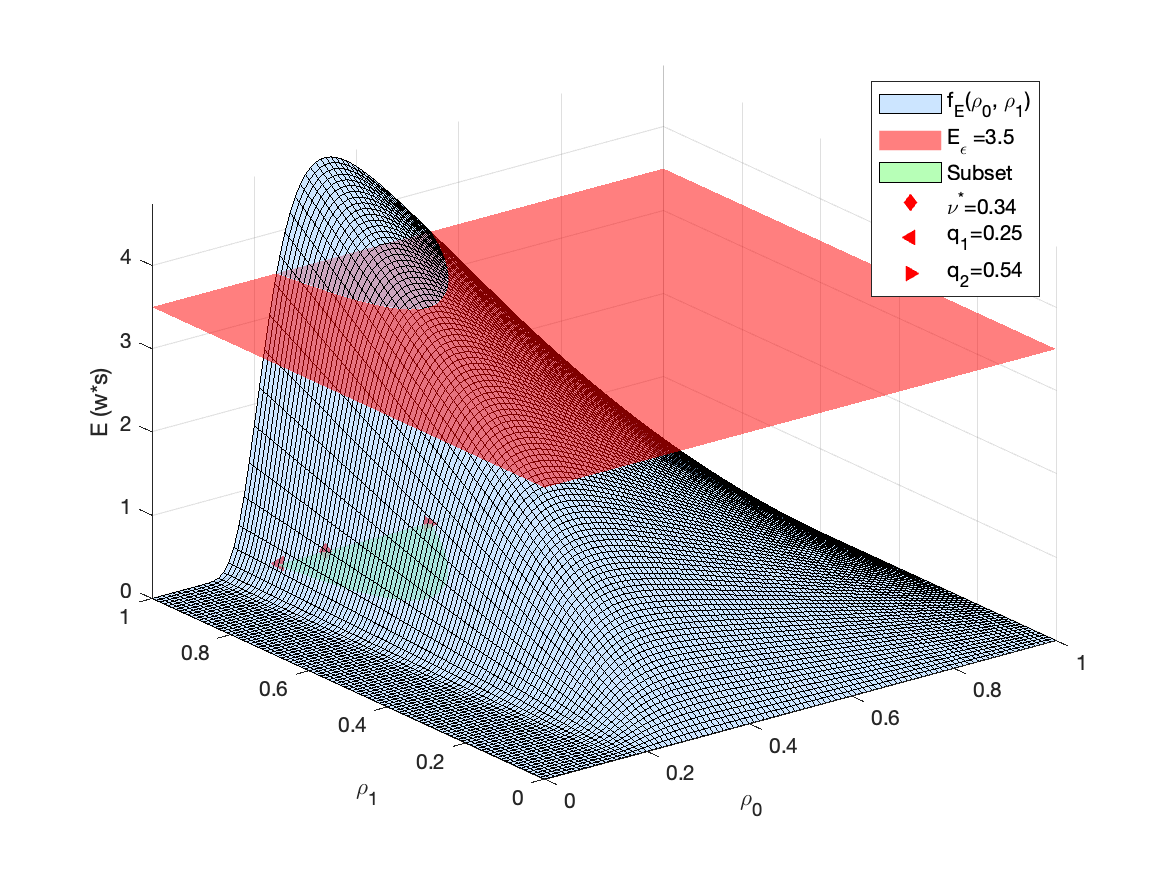}
        \caption*{(a) $E \geq E_{\epsilon} = 3.5 \ w*s$.} 
    \end{minipage}
    \hspace{0.5cm} 
    \begin{minipage}{0.45\textwidth}
        \centering
        \includegraphics[width=1\linewidth]{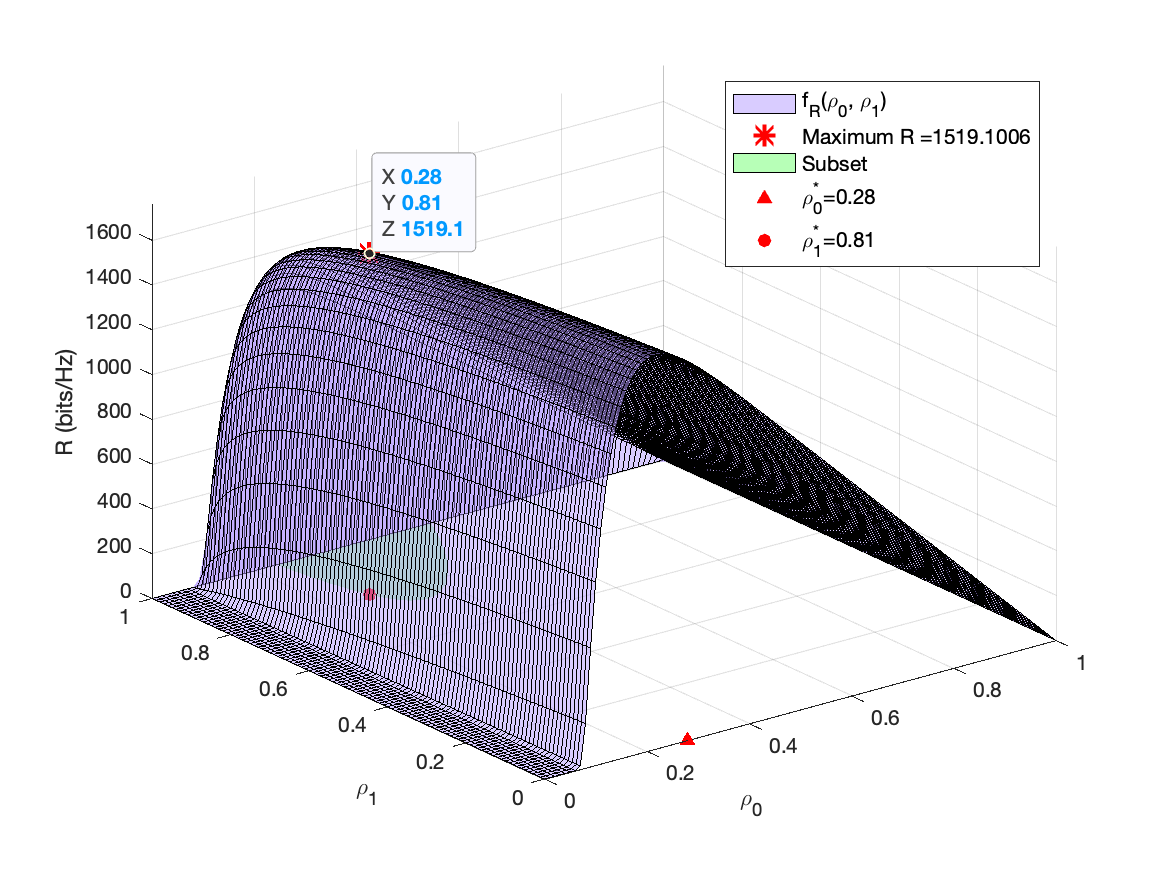} 
        \caption*{(b) Maximum value of $R$.} 
    \end{minipage}
    \caption{Maximize $R$ with $E \geq E_{\epsilon}$.}
    \label{fig-rho_01_MaxCfixE}
\end{figure}
Fig. \ref{fig-rho_01_MaxCfixE} shows the maximization of $R$ with respect to $\rho_0$ and $\rho_1$ for $E \geq E_{\epsilon}$. In this figure, $E_{\epsilon}$ is set to 3.5 $(w*s)$, and other parameters are the same as in Fig. \ref{fig-rho_01_MaxEfixC}. Similarly, Fig. \ref{fig-rho_01_MaxCfixE} (a) shows that $E$ changes with $\rho_0$ and $\rho_1$. Meanwhile, the plane shows the locations of $E_{\epsilon} = 3.5 (w*s)$, by which a subset of $\mathcal{D}$ that meets $E \geq E_{\epsilon}$ can be obtained as shown in the shadow region in $\rho_0\textendash\rho_1$ plane according to  Algorithm \ref{alg-P2}. In this case, the values of $\nu$, $q_1$ and $q_2$ are obtained as $0.24$, $0.25$ and $0.54$, respectively, as indicated by the diamond, the left triangle and the right triangle in the figure. Subsequently, Fig. \ref{fig-rho_01_MaxCfixE} (b) shows the maximum value of $R$ on this subset. One sees that the $R^{*}$ is 1519.1 $(bits/Hz)$, and the corresponding ${\rho_0}^*$ and ${\rho_1}^*$ are $0.28$ and $0.81$, respectively, as indicated by the upper triangle and the solid circle in the figure. Note that the optimal values of $\rho_0$ in both Fig. \ref{fig-rho_01_MaxEfixC} and Fig. \ref{fig-rho_01_MaxCfixE} are the same. This means that when the sensing time ratio is set to ${\rho_0}^* = 0.28$ in the first phase, the best performance of communications \& powering in the second phase can be balanced in this case.


\begin{figure}[t]
\centering
\includegraphics[width=1\linewidth]{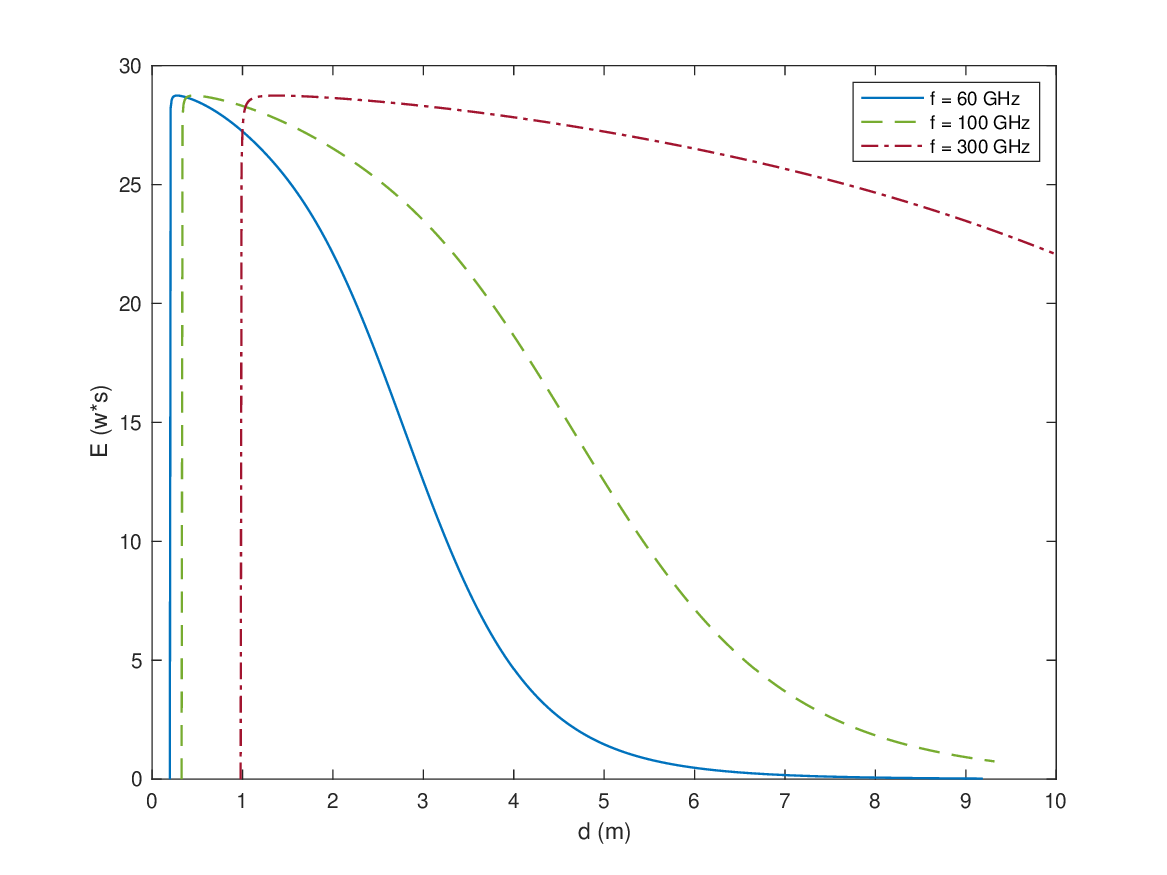}
\caption{$E$ vs $d$}
\label{fig-requency123_d}
\end{figure}
Fig. \ref{fig-requency123_d} shows the harvested energy $E$ versus $d$ when $f$ is set to 60 $GHz$, 100 $GHz$ and 300 $GHz$. In the figure, $\rho_0 = 0.4$ and $\rho_1 = 1$. Generally speaking, the total received energy increases quickly and then decreases with $d$. For different operating frequencies, one can see that the minimum distance, i.e., $d_{min}$ in \eqref{eqns-Near-Field-Gamma-Minimum-d}, increases. This is because a higher $f$ leads to a shorter $\lambda$, and a larger $\lambda G_{bs-t} = \eta_{bs-t}(\pi {D_{bs-t}})^{2}/\lambda$ in \eqref{eqns-Near-Field-Gamma-Minimum-d}. In addition, the Rayleigh distance $d_R = 2{D_{bs-t}}^{2}/\lambda$ also becomes longer with shorter $\lambda$, which means the FZ region becomes wider. When $d$ is in FZ region, on the one hand, $\gamma_A$ increases with $d$, which makes the path loss factor $PL(d)$ larger. On the other hand, a longer $d$ also causes more atmospheric absorption, which in turn makes $PL(d)$ smaller. As a result, there is a threshold within the FZ region that maximizes the received total energy. When $d$ is less than this threshold, the received total energy increases with $d$, and when $d$ exceeds the threshold, the received total energy decreases with $d$. This is why the received energy increases rapidly to its maximum and then decreases. Besides, It can also be observed that, as the frequency $f$ increases, the total received energy decreases more slowly with $d$. This is because a higher $f$ leads to a wider FZ region.

\begin{figure}[t]
\centering
\includegraphics[width=1\linewidth]{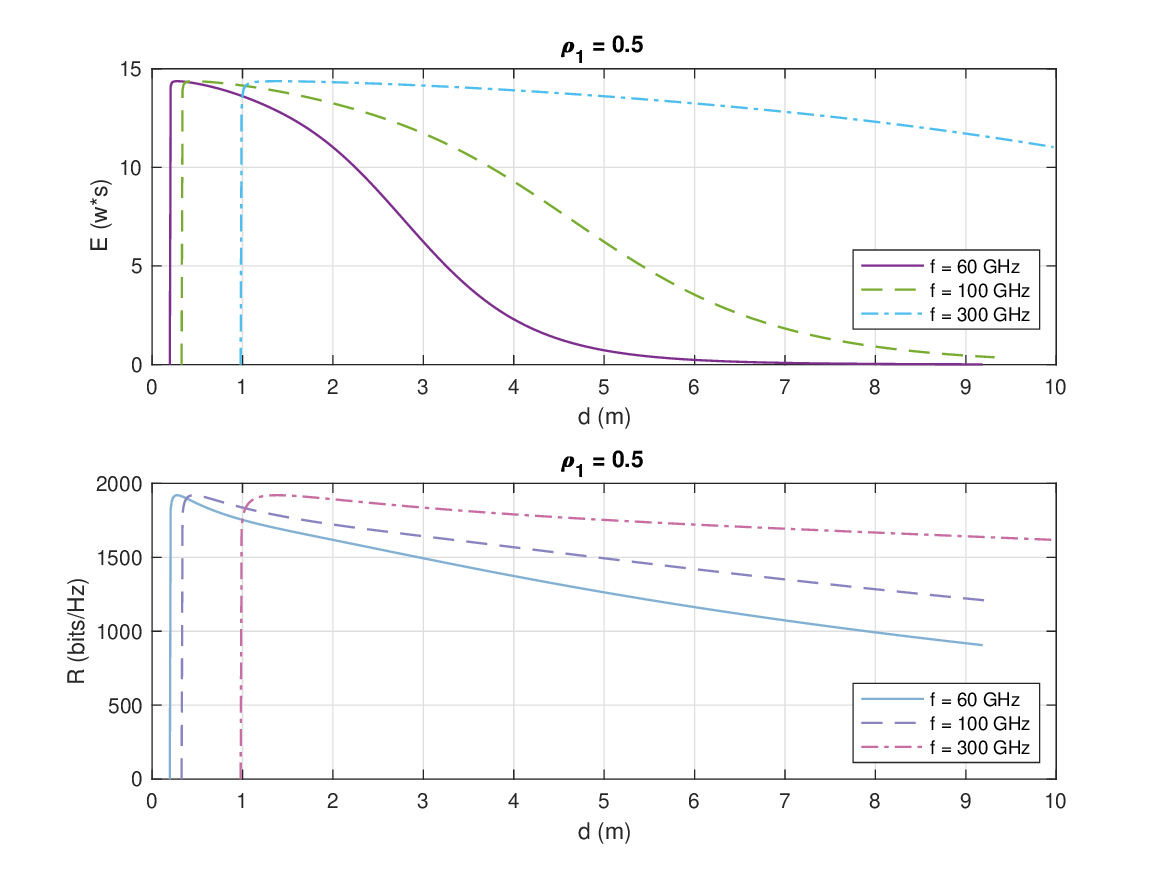}
\caption{$E$ and $R$ vs $d$.}
\label{frequency123_d_EandC}
\end{figure}
Fig. \ref{frequency123_d_EandC} shows the harvested energy $E$ and achievable rate $R$ versus $d$. In this figure, the parameter settings are the same as those in Fig. \ref{fig-requency123_d}, except that $\rho_1$ is fixed to 0.5. From both parts of the figure, the same observations can be made as in Fig. \ref{fig-requency123_d}. Again, with a higher $f$, $d_{min}$ becomes larger and the FZ region becomes wider.

\begin{figure}[t]
\centering
\includegraphics[width=1\linewidth]{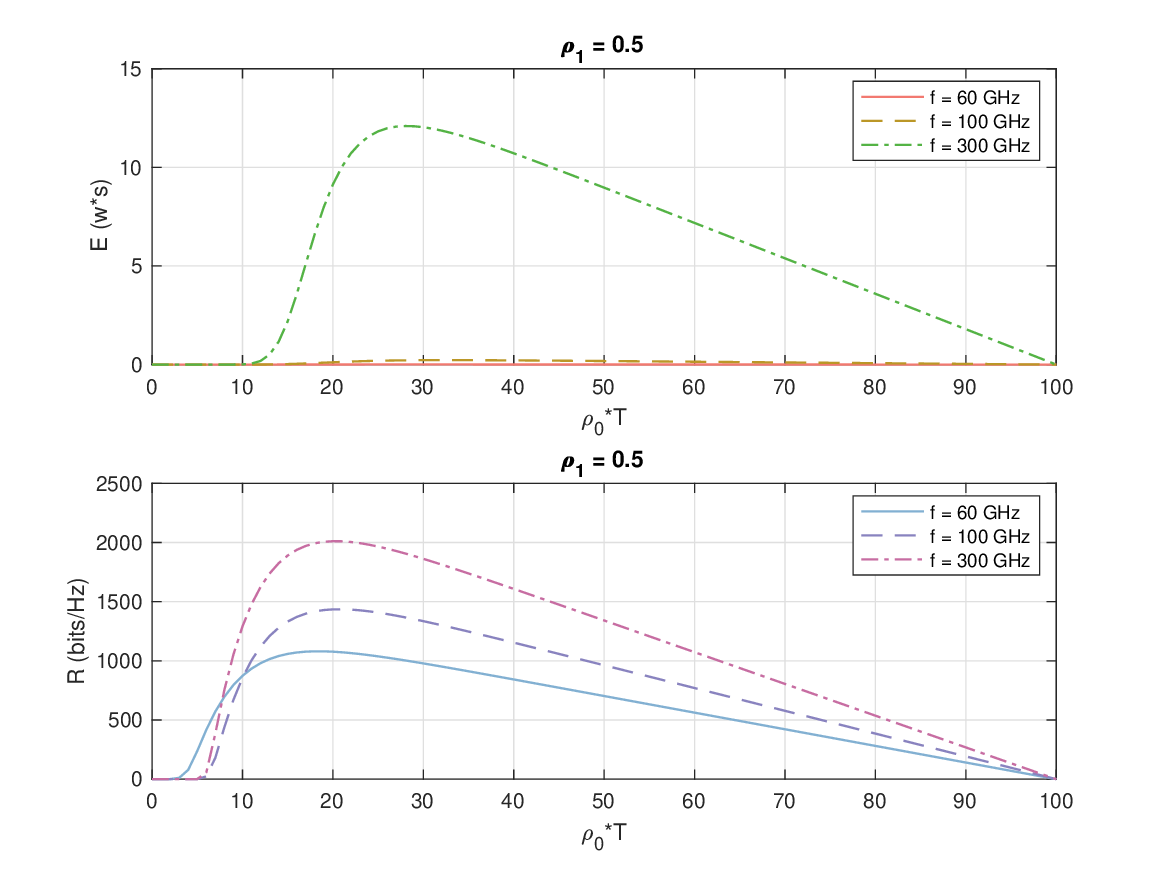}
\caption{$E$ and $R$ vs $\rho_0T$.}
\label{fig-frequency123_T_EandC}
\end{figure}
Fig. \ref{fig-frequency123_T_EandC} compares the impact of sensing time on the harvested energy $E$ and the achievable rate $R$ for frequencies of 60 $GHz$, 100 $GHz$ and 300 $GHz$. In this figure, $d = 10\ m$ and $\rho_1 = 0.5$. From the figure, one can see that both harvested energy and achievable rate increase first and reach its maximum, and then they decrease to 0 with $\rho_0$. This agrees with the observations from Fig. \ref{fig-Energy_Communications}. Particularly, for those three different frequencies, one sees that the higher the frequency is, the greater the harvested energy and achievable rate will be. This is because although increasing frequency results in greater path loss and more molecular absorption, a higher gain and a greater beam collection efficiency $\eta_b$ associated with higher frequencies could compensate for these losses, leading to higher received power. Consequently, it leads to increased harvested energy and improved achievable rate. Next, we will further investigate the effect of frequency on the harvested energy and achievable rate.

\begin{figure}[t]
\centering
\includegraphics[width=1\linewidth]{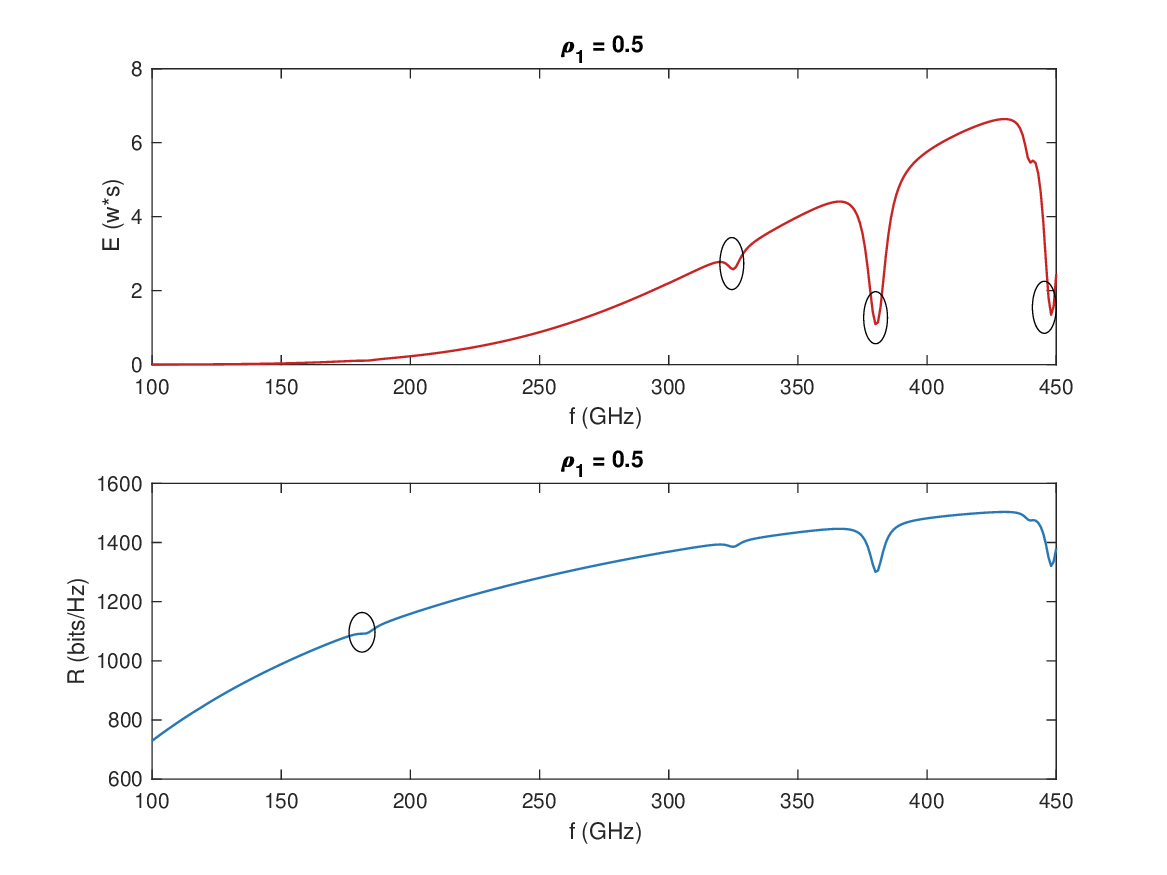}
\caption{$E$ and $R$ vs $f$.}
\label{fig-frequencies_EandC}
\end{figure}
Fig. \ref{fig-frequencies_EandC} examines the effect of frequency on the harvested energy and achievable rate. In this figure, $\rho_0 = 0.4$, $\rho_1 = 0.5$, $d = 20\ m$ and $f$ changes from 100 $GHz$ to 450 $GHz$. From the figure, one can see that both harvested energy and achievable rate show an upward trend with increasing $f$, which further validates our analysis in Fig. \ref{fig-frequency123_T_EandC}. However, as indicated by the ellipses in the upper part of the figure, corresponding to frequencies around 184 $GHz$, 326 $GHz$, 381 $GHz$ and 449 $GHz$, both the harvested energy and achievable rate show an abnormal glitch. This occurs because frequencies in these ranges experience significantly higher losses \cite[Fig. 3 and Fig. 8]{Kokkoniemi2021}. Based on this observation, careful consideration of frequency selection is of great importance in the design of THz-ISCAP systems.

\begin{figure}[t]
\centering
\includegraphics[width=1\linewidth]{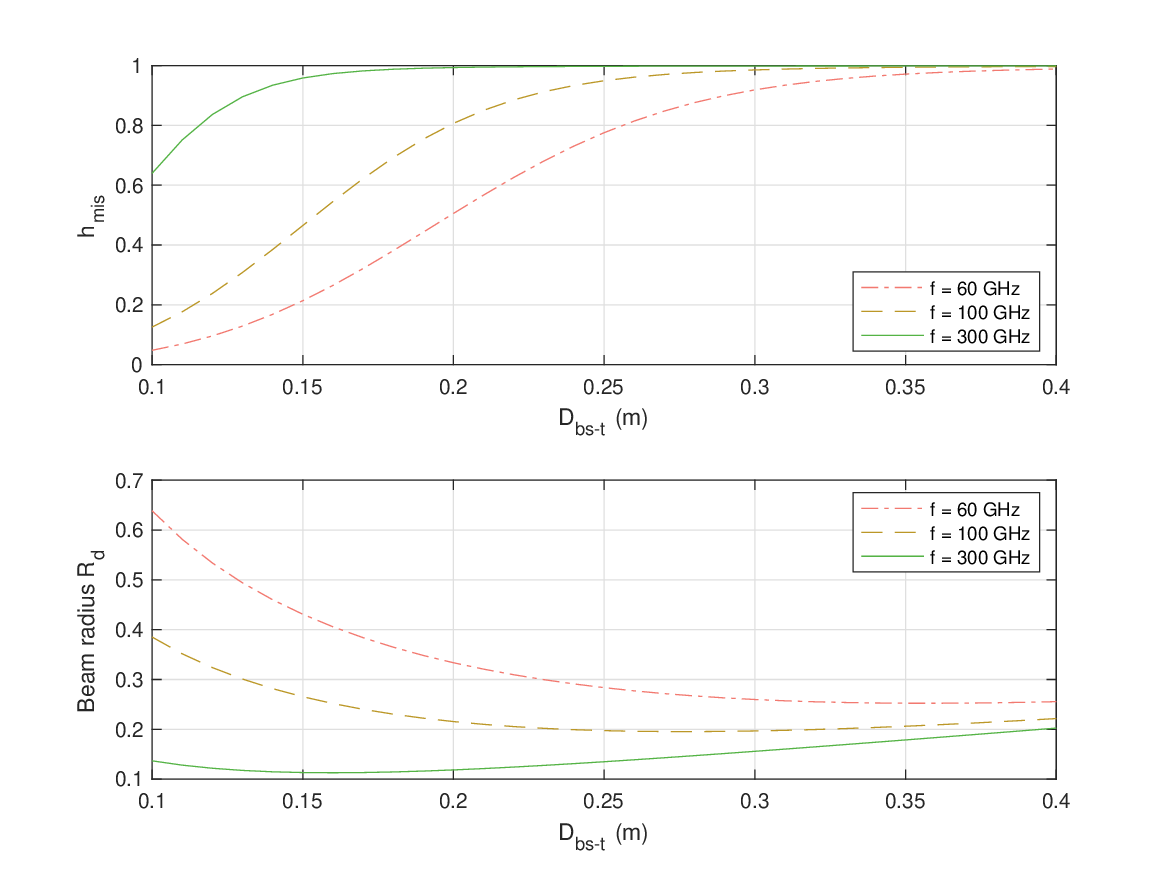}
\caption{$h_{mis}$ and $R_d$ vs $D_{bs-t}$.}
\label{fig-frquency123_Dbs_hmis_bw}
\end{figure}
Fig. \ref{fig-frquency123_Dbs_hmis_bw} investigates the effect of aperture diameter on misalignment coefficient and beam radius in THz-ISCAP. In the figure, the aperture diameters of the BS and the receiver are set in the ranges of 0.1 $m$ to 0.4 $m$ and 0.2 $m$ to $0.8$ m, $\rho_0 = 0.4$, $\rho_1 = 0.5$, and $d = 20\ m$. In the upper part of the figure, one can see that $h_{mis}$ increases with $D_{bs-t}$. This is because when increasing the aperture diameters of both antennas, the beam radius $R_d$ decreases first and then increases, as shown in the lower part of the figure. This leads to an increased $\epsilon$ and $S_0$ in \eqref{eqn-S_0} and hence, leading to an increased $h_{mis}$. For $R_d$ in the lower part of the figure, according to \eqref{eqn-Gaussian-Beams}, the beam waist $W_0$ is determined by the aperture diameter, and it increases with the aperture diameter. Given $f$, $(d/d_0)^2$ in \eqref{eqn-Gaussian-Beams} decreases with $D_{bs-t}$. However, the decrease in $(d/d_0)^2$ compared to the increase of $W_0$ can be ignored as $D_{bs-t}$ increases, and this is why $R_d$ decreases first and then increases. In addition, for a larger $f$, $\lambda$ is smaller, which makes $R_d$ quickly reach its minimum and then increase. This is why the value of $h_{mis}$ with high frequency (i.e., $f = 300\ GHz$) is higher than that with low frequency (i.e., $f = 100\ GHz$) at the beginning.

\begin{figure}[t]
\centering
\includegraphics[width=1\linewidth]{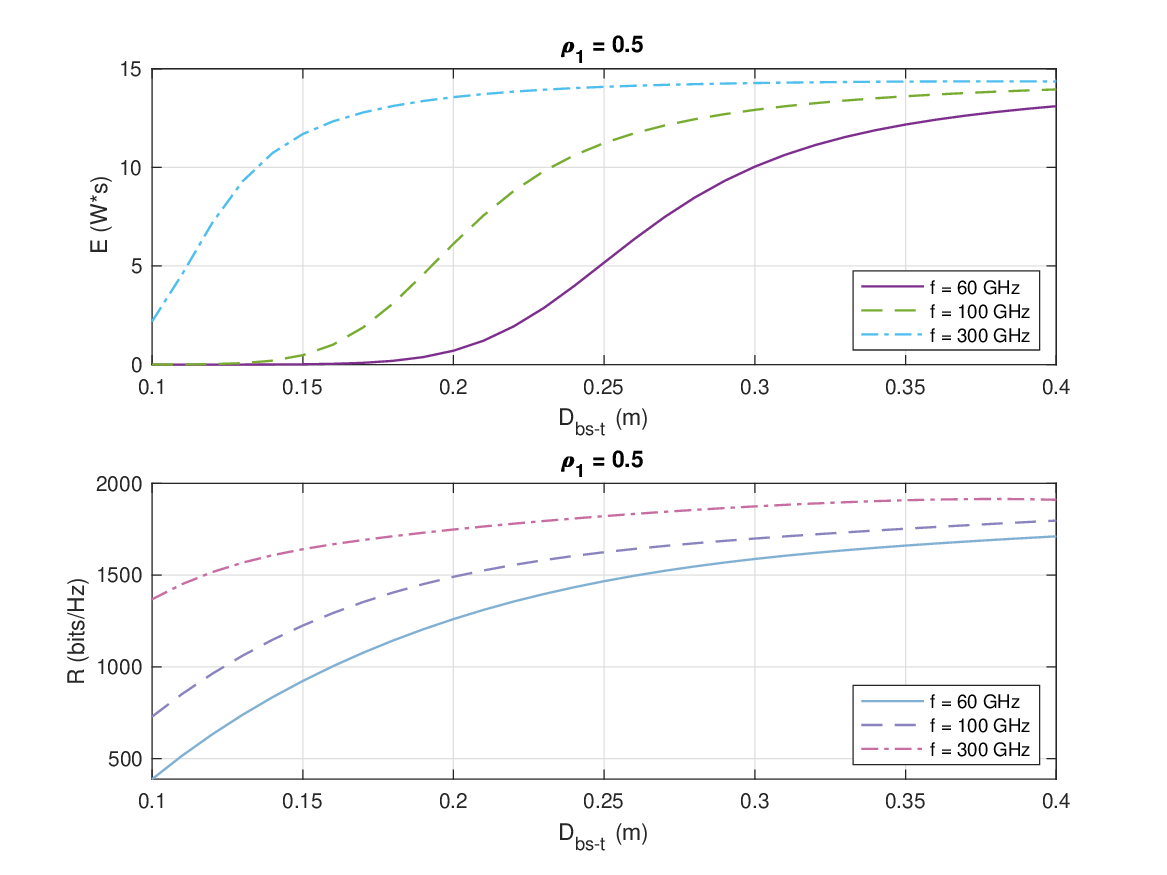}
\caption{$E$ and $R$ vs $D_{bs-t}$.}
\label{fig-frquency123_Dbs_EandC}
\end{figure}
Fig. \ref{fig-frquency123_Dbs_EandC} shows the harvested energy and achievable data versus $D_{bs-t}$. Using the same parameter settings as those in Fig. \ref{fig-frquency123_Dbs_hmis_bw}, one can see from the figure that both $E$ and $R$ increase with $D_{bs-t}$.This is because, with a higher $h_{mis}$, the received power will be higher, leading to a higher $E$ and $R$. From Fig. \ref{fig-frquency123_Dbs_hmis_bw}, the higher the $f$ is, the greater the $h_{mis}$ will be. As a result, with a higher $f$, $E$ and $R$ will be higher, as shown in the figure.

\section{Conclusions}
\label{Conclusions}
In this paper, a THz-ISCAP system where sensing has been used for improving the performance of communications and powering by reducing misalignment has been studied. The simulation results have shown that, within a fixed time interval $T$, there is an optimal time ratio $\rho_0$ for sensing to reduce the misalignment so as to improve the performance of communications and powering. Meanwhile, an optimal power splitting ratio $\rho_1$ is also derived to maximize the achievable data rate or received energy by meeting a minimum requirement on the other. These findings provide new insights into the system design of sensing-assisted THz-SWIPT.



%
%

\end{document}